\RequirePackage{rotating}
\documentclass[acmsmall]{acmart}

\def\BibTeX{{\rm B\kern-.05em{\sc i\kern-.025em b}\kern-.08emT\kern-.1667em\lower.7ex\hbox{E}\kern-.125emX}}

\usepackage[ruled]{algorithm2e} % For algorithms

\SetAlFnt{\small}
\SetAlCapFnt{\small}
\SetAlCapNameFnt{\small}
\SetAlCapHSkip{0pt}
\IncMargin{-\parindent}

\acmJournal{CSUR}

\usepackage{amsmath, amssymb}
\usepackage{makecell}
\usepackage{multirow}
\usepackage{multicol}
\usepackage{url}
\usepackage[T1]{fontenc}

\definecolor{mediumaquamarine}{rgb}{0.4, 0.8, 0.67}
\definecolor{bittersweet}{rgb}{1.0, 0.44, 0.37}
\definecolor{sandybrown}{rgb}{0.96, 0.64, 0.38}
\definecolor{saffron}{rgb}{0.96, 0.77, 0.19}
\definecolor{salmonpink}{rgb}{1.0, 0.57, 0.64}
\definecolor{limegreen}{rgb}{0.2, 0.8, 0.2}
\definecolor{persianorange}{rgb}{0.85, 0.56, 0.35}
\definecolor{darkgray}{rgb}{0.66, 0.66, 0.66}

\usepackage{tabularx}
\usepackage{color, colortbl}
\definecolor{Gray}{gray}{0.9}
\usepackage{hhline}
\usepackage{rotating}
\usepackage{lscape}
\usepackage[section]{placeins}
\usepackage{longtable}
\usepackage{tablefootnote}
\usepackage{graphicx}
\begin{document}

\title{Metaverse: Survey, Applications, Security, and Opportunities}

\author{Jiayi Sun}
\affiliation{
	\institution{Jinan University}
	\postcode{510632}
	\city{Guangzhou}
	\state{Guangdong}
	\country{China}}
\email{jiayisun01@gmail.com}

\author{Wensheng Gan*}
\affiliation{
	\institution{Jinan University}
	\postcode{510632}
	\city{Guangzhou}
	\state{Guangdong}
	\country{China}}
\email{wsgan001@gmail.com}

\author{Han-Chieh Chao}
\affiliation{
	\institution{National Dong Hwa University}
	\postcode{974301}
	\city{Hualien}
	\state{Taiwan}
	\country{R.O.C.}}
\email{hcc@ndhu.edu.tw}

\author{Philip S. Yu}
\affiliation{
	\institution{University of Illinois at Chicago}
	\postcode{60607}
	\city{Chicago}
	\state{IL}
	\country{USA}}
\email{psyu@uic.edu}

% By default, the full list of authors will be used in the page headers. Often, this list is too long, and will overlap
	% other information printed in the page headers. This command allows the author to define a more concise list
	% of authors' names for this purpose.
\renewcommand{\shortauthors}{Jiayi Sun et al.}
	
	%
	% The abstract is a short summary of the work to be presented in the article.
	
\begin{abstract}
	As a fusion of various emerging digital technologies, the Metaverse aims to build a virtual shared digital space. It is closely related to extended reality, digital twin, blockchain, and other technologies. Its goal is to build a digital space based on the real world, form a virtual economic system, and expand the space of human activities, which injects new vitality into the social, economic, and other fields. In this article, we make the following contributions. We first introduce the basic concepts such as the development process, definition, and characteristics of the Metaverse. After that, we analyze the overall framework and supporting technologies of the Metaverse and summarize the status of common fields. Finally, we focus on the security and privacy issues that exist in the Metaverse, and give corresponding solutions or directions. We also discuss the challenges and possible research directions of the Metaverse. We believe this comprehensive review can provide an overview of the Metaverse and research directions for some multidisciplinary studies.
\end{abstract}
	

%	\terms{Education, Blockchain}
	

\maketitle

\section{Introduction}
\label{sec:introduction}

\textbf{Background.} The advent of the computer in 1946 freed the human brain from tedious calculations. With the continuous reduction of the size of the computer, it is becoming increasingly popular in people's lives. The birth of the Internet in 1969 connected individual computers and brought new ways of information exchange for people. While expanding and improving the means and capabilities of mass communication, computers and the Internet have greatly promoted social development and the Cultural Revolution. The form of people's access to the Internet is also constantly evolving, starting with the PC Internet at the beginning and then moving to the mobile Internet. From another perspective, Internet is being integrated into people's lives to a greater extent and at a more granular level. Before the advent of Internet, information was overwhelmingly produced by human society. Internet has enabled the interconnection of computers and various devices, and various types of information have emerged in large quantities. In the past few years, non-human societies in physical space (such as sensors) have also generated a lot of information. This information blend and influence each other, and gradually form the third space outside human society and physical society -- cyberspace\footnote{https://en.wikipedia.org/wiki/Cyberspace}.

Cyberspace is a new dimension that is growing rapidly. The interaction and development of the ternary space has prompted mankind to enter the era of big data. With the continuous extension of the network, the corresponding digital technologies are also advancing, such as big data, artificial intelligence, and virtual reality. The creation of these technologies bridged the real world and the digital world, and the real lives of human beings began to migrate to the virtual world on a large scale. The emergence of isolation approaches caused by COVID-19 has led to a significant increase in the online time of the whole society. Digital life has changed from a temporary exception to the norm, which is gradually becoming the parallel world of the real world instead of the supplement of the real world. People are no longer satisfied with the flat screen. The three-dimensional interaction of three-dimensional space has become a real need. The real-time interaction of the ternary space requires a high-speed, low-latency, stable, and reliable large-scale communication environment, massive data processing, cloud real-time rendering, and powerful intelligent computing. It is the maturity of virtual reality (VR) \cite{slater2018immersion,hudson2019or,hu2021virtual}, artificial intelligence (AI) \cite{zhang2021study,huynh2022artificial}, blockchain (BC)\footnote{https://en.wikipedia.org/wiki/Blockchain} \cite{swan2015blockchain}, digital twin (DT) \cite{tao2018digital,ibm2021dt}, 5G \cite{shafique2020internet}, Web 3.0\footnote{https://en.wikipedia.org/wiki/Web3}, and other technologies that make it possible to realize this illusory vision now. The emergence and convergence of various digital technologies in recent years have brought the concept of the Metaverse into focus, making it an attractive vision.

\textbf{Research status.} There is no clear definition of the Metaverse. Researchers have provided insights with their own focus, but their most essential connotation is consistent. The Metaverse is a highly immersive virtual digital world formed by digital technology in which people can simulate various activities in the real world and interact with the real world. It is the next generation of Internet applications and social forms resulting from the integration of multiple new technologies, making the ternary world more inseparable. The term Metaverse was first coined by Neal Stephenson in 1992 \cite{stephenson2003snow}. Since then, more and more visions of the Metaverse appeared in movies, games, and other cultural works. Second Life\footnote{https://secondlife.com}, an online virtual game developed by Linden LABS in 2003, was the first phenomenally virtual world. In the game, players interact with each other in the form of virtual avatars to complete social activities such as entertainment, production, and trading. In 2006, Roblox\footnote{https://www.roblox.com}, a game with a focus on virtual worlds and self-built content, was launched. Currently, it is the world's largest multiplayer online creation platform.

With the rapid development of digital technologies and related bioinformatics, the implementation of the Metaverse is gradually becoming a reality. In 2020, the COVID-19 outbreak led to a rapid increase in the demand for contactless social activities, which greatly promoted social virtualization. In 2021, the attention of the Metaverse skyrocketed. In March 2021, Roblox, an American company known as "the first share of Metaverse", went public and gave great attention to the concept of the Metaverse for the first time. As the big companies continue to follow up, the popularity of the Metaverse continues to soar. The emergence of the Metaverse is not sudden; it is an inevitable product of the integration of technologies in various fields. However, the Metaverse is in its early stages. In other words, there are still many difficulties and obstacles to be overcome in the realization of the Metaverse. The first is the technical implementation. The point where the various technologies come together in the Metaverse is something that needs to be carefully considered, including the inherent flaws that the technologies do not address and the new problems that come with the fusion \cite{yang2022fusing}. Next is a series of problems and impacts caused by the application of the Metaverse to the real world. There are many potential problems associated with the interaction between the virtual and real worlds, such as economic development and social culture. Finally, the emergence of highly influential new technologies will inevitably change social patterns, thus triggering thoughts about new social systems. In addition, the privacy and security of the Metaverse are some issues that must be guaranteed.

\begin{figure*}[h]
	\centering
	\includegraphics[scale = 0.39]{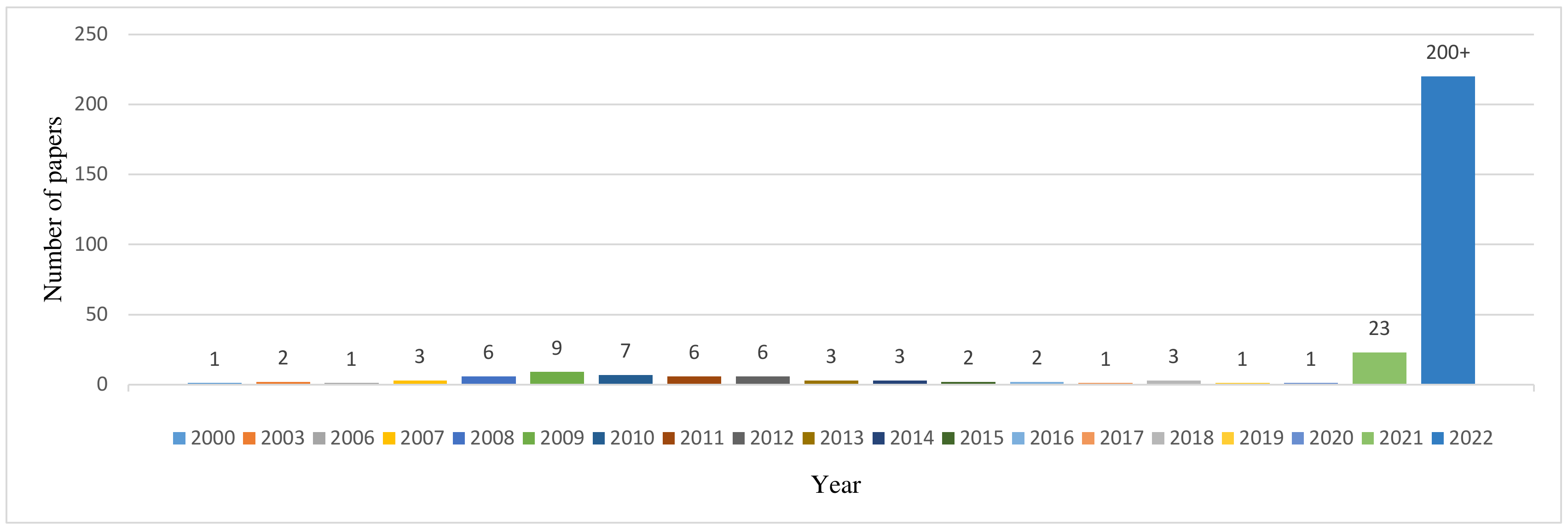}
	\caption{Statistics on the number of papers related to the Metaverse}
	\label{fig:Papernumber}
\end{figure*}

\textbf{Paper statistics.} The year-wise articles of the Metaverse selection statistics are depicted in Fig. \ref{fig:Papernumber}. It is clear that there exists a strong demand for the development and application of the Metaverse. With the breakthroughs of the Internet of Things (IoT) \cite{stoyanova2020survey,nguyen20216g}, extended reality (XR) \cite{ratcliffe2021extended,rauschnabel2021augmented,xi2022challenges}, and other related technologies, the Metaverse has flourished in recent years, especially since 2021.

%%%%%%%%%%%%%%%%%%   number of paper    %%%%%%%%%%%%%%%%%%
%%%%%%%%%%%%%%%%%%   number of paper   %%%%%%%%%%%%%%%%%%%

\textbf{Research gaps.} Note that the scope of this survey includes the Metaverse and its supporting technologies and applications, as well as security, privacy, and other fields. The implementation and application of the Metaverse are increasingly active areas of research and the industrial community, with a lot of existing research and opportunities in the recent literature. There are some surveys that introduce frameworks, technologies, or security related to the Metaverse separately or simultaneously, but they do not summarize the current status and prospects of the Metaverse projects in common areas comprehensively. For example, some surveys about the technical framework and existing problems of the Metaverse can be found in \cite{ning2021survey,lee2021all,park2022metaverse}. Ning \textit{et al.} \cite{ning2021survey} summarized the Metaverse-related policies and application status of various countries, but did not pay attention to specific security problems and solutions. There are also reviews that focus on the adaptations of individual technologies in the Metaverse. For example, some surveys related to the fusion of AI or BC with the Metaverse and their role in the virtual world are covered in \cite{huynh2022artificial,yang2022fusing,gadekallu2022blockchain}. XR and related applications in the Metaverse are introduced in \cite{mystakidis2022metaverse}. While none of them pays attention to the current state of Metaverse projects in various fields, and provides no solution to the Metaverse privacy problems. There are some reviews that mainly summarize the security and privacy issues of the Metaverse, such as \cite{wang2022survey,zhao2022metaverse}, which includes security issues and corresponding solutions. However, there has been no survey devoted to the discussion ranging from the technical framework of the Metaverse to security and privacy issues and the prospect of Metaverse applications in major domains. This motivates us to provide an overview of Metaverse-related concepts, techniques, security, and applications. In summary, the contributions and gaps of the existing surveys are listed separately in Table \ref{table:Gaps}.

%Gaps in Existing Surveys
\begin{table}
	\centering
	\caption{Contributions and gaps of existing surveys}
	\scalebox{0.8}{
		\begin{tabular}{|c|c|p{5cm}|c|c|c|}
			\hline
			\textbf{Ref.} & \textbf{Year}   & \textbf{One-sentence summary}  & \textbf{Applications} & \textbf{Security threats} & \textbf{Security solutions} \\ \hline
			
			\cite{huynh2022artificial}    & 2022   &  Artificial intelligence for the Metaverse: A Survey   & \checkmark   & $\times$ &  \checkmark          \\ \hline
			
			\cite{yang2022fusing} & 2022   & Fusing blockchain and AI with Metaverse: A survey & $\times$   & \checkmark & $\times$         \\ \hline
			
			\cite{lee2021creators} & 2021  & When creators meet the Metaverse: A survey on computational arts & \checkmark   & $\times$
			&  $\times$          \\ \hline
			
			\cite{ning2021survey}   & 2021  & A Survey on Metaverse: The state-of-the-art, technologies, applications, and challenges  & \checkmark   & $\times$ &  \checkmark         \\ \hline
			
			\cite{lee2021all}  & 2021  & A complete survey on technological singularity, virtual ecosystem, and research agenda  & $\times$  & \checkmark &  \checkmark          \\ \hline
			
			\cite{wang2022survey}& 2022   & A survey on Metaverse: Fundamentals, security, and privacy & $\times$   & \checkmark &  \checkmark          \\ \hline
			
			\cite{gadekallu2022blockchain}   & 2022   &  Blockchain for the Metaverse: A Review   & $\times$   & \checkmark &  \checkmark               \\ \hline
			
			\cite{jeon2022blockchain}    & 2022    & Blockchain and AI meet in the Metaverse   & $\times$   & \checkmark &  \checkmark               \\ \hline
			
			\cite{park2022metaverse}    & 2022   &  A Metaverse: Taxonomy, components, applications, and open challenges  & \checkmark   & \checkmark &  $\times$               \\ \hline
			
			\cite{zhao2022metaverse}  & 2022    &  Metaverse: Security and privacy concerns    & $\times$   & \checkmark  & \checkmark                \\ \hline
			
			\cite{mystakidis2022metaverse}  & 2022    &  Contemporary Metaverse development, Meta-education and innovative applications    & \checkmark   & $\times$  & $\times$                \\ \hline
			
			Our work & 2022  & Metaverse: Technologies, applications, security, and opportunities  & \checkmark&  \checkmark & \checkmark\\ \hline
		\end{tabular}
	}
	\label{table:Gaps}
\end{table}

\textbf{Contributions.} This article conducts a comprehensive survey of the Metaverse concept, research status, technical framework, etc., and presents existing problems and solutions for its security and privacy. To the best of our knowledge, this is the first paper that reviews in detail the definitions, technical frameworks, features, applications, security and privacy, threats and solutions of the Metaverse. In light of previous studies and surveys, our survey aims to (i) provide a presentation of various definitions and frameworks of the Metaverse; (ii) give an in-depth discussion of the applications of the Metaverse in various fields; and (iii) review the security and privacy issues in the Metaverse and summarize possible solutions. In summary, the major contributions of this paper are highlighted as follows:

\begin{enumerate}
	\item[(1)] Firstly, we summarized the different definitions of the Metaverse and various reasonable technical frameworks, and put forward the characteristics of the Metaverse.
	\item[(2)] We are the first to provide a systematic overview of the Metaverse projects in various fields and future applications, including: education, medicine, smart city, business, culture, manufacturing, and others.
	\item[(3)] We summarize the threats to security and privacy in the Metaverse in detail, and then provide an in-depth discussion for the corresponding solutions.
	\item[(4)] Finally, we focus on the technical challenges and future directions of the Metaverse.
\end{enumerate}

\textbf{Roadmap.} This article is organized as follows. We introduce the concepts and characteristics of the Metaverse in Section 2. In Section 3, we describe the core technologies and framework of the Metaverse and explain how its ecosystem is constructed. In Section 4, we summarize the Metaverse projects, including specific applications in education, health care, business, and other fields. In Section 5, we discuss the security problems faced by the Metaverse in its current development stage and possible solutions. In Section 6, we look forward to the future direction of Metaverse and the challenges it may encounter. The Metaverse, formed by the fusion of various digital technologies, will greatly promote social development. We summarize the paper in Section 7. Table \ref{tab:abbreviations} contains a list of abbreviations that appear frequently in this article.

\begin{table}[h]
	\centering
	\caption{List of abbreviations in alphabetical order.}
	\label{tab:abbreviations}
	\scalebox{0.76}{
		\begin{tabular}{|c|c|}
			\hline
			\textbf{Acronym} &  \textbf{Explanation} \\\hline\hline
			5G & 5\textsuperscript{th} Generation Mobile Networks \\\hline	
			AI & Artificial Intelligence   \\\hline
			BC & Blockchain   \\\hline
			BCI & Brain-Computer Interface \\\hline
			VR & Virtual Reality   \\\hline 
			AR & Augmented Reality  \\\hline
			AIGC & Artificial Intelligence Generated Content \\\hline
			MR & Mixed Reality  \\\hline
			XR & Extended Reality   \\\hline	
			DT & Digital Twin \\\hline
			DAO  & Decentralized Autonomous Organization \\\hline
			IoT & Internet of Things \\\hline

			MUD & Multi-User Dungeon \\\hline
			MMORPG & Massively Multiplayer Online Role-Playing Game \\\hline
			MMOG & Massively Multiplayer Online Game \\\hline
			
			PGC & Professionally-Generated Content \\\hline				
			UGC & User-Generated Content \\\hline
			NFT & Non-Fungible Token  \\\hline
			NLP & Natural Language Processing \\\hline
			NPC & Non-Player Character\\\hline
			
		\end{tabular}
	}
\end{table}

%%%%%%%%%%%%%%%%%%%%%%%%%%%%% separator %%%%%%%%%%%%%%%%%%%%%%%%%%%%%%%%

\section{Metaverse} \label{sec:Metaverse}

\subsection{The timeline of the Metaverse}

\textbf{Incubation.} In 1979, the first Multi-User Dungeon (MUD)\footnote{https://en.wikipedia.org/wiki/MUD} appeared. It uses a text interface to connect multiple users together in real time, forming an open social and cooperative world. In 1986, Habitat, the world's first MUD online game with a 2D graphical interface, was launched. It is considered the forerunner of modern massively multiplayer online role-playing games (MMORPGs)\footnote{https://en.wikipedia.org/wiki/Massively\_multiplayer\_online\_role-playing\_game}, allowing humans to enter virtual worlds for the first time using avatars. In 1984, Canadian writer William Ford Gibson coined the word "cyberspace" in his science fiction novel Neuromancer, which also gave rise to the "cyberpunk" culture. The novel depicts the rich virtual space behind the computer screen, giving people an advanced look into the digital world of the future. The concept of "Metaverse" and "avatar" was first explicitly introduced in 1992 by Neil Stephenson in his novel Snow Crash \cite{stephenson2003snow}. In the novel, a real human and a virtual human live together in a virtual space through VR equipment, showing the space-time extension and man-machine symbiosis of the Metaverse. The Metaverse has been on the scene ever since.

\textbf{Preparation.} The most likely starting point for the Metaverse implementation is the gaming sector. In 1994, Web World, the first multiplayer social game with an axonometric interface, was introduced. In this game, users can socialize in real time and transform the game world. The user generated content (UGC)\footnote{https://en.wikipedia.org/wiki/User-generated\_content} mode was enabled. In 1995, Worlds Inc.\footnote{https://www.worlds.com/} became the first massively multiplayer online game (MMOG or more commonly MMO)\footnote{https://en.wikipedia.org/wiki/Massively\_multiplayer\_online\_game} to market, hoping that users could conduct open social interactions in 3D space. In 1995, based on the novel Snow Crash, Active Worlds\footnote{https://www.activeworlds.com/} was created. Active Worlds provides users with basic content creation tools to personalize 3D virtual environments. In 2003, Linden Laboratory in the United States developed the phenomenal-level virtual world "Second Life", which has powerful world editing functions. Unlike previous digital games, this game has a virtual economy that generates value in the real world and can be used in areas such as education. Therefore, a large number of enterprises and institutions were attracted. In 2006, the multiplayer online creation game Roblox was officially released. With a rudimentary version of the Metaverse, the majority of the works in the game are user-created, interactive, and linked to real life. Due to the COVID-19 pandemic, people's social lives are gradually moving online. The rapid growth of the demand for online interaction accelerates the maturity of emerging technologies such as XR, digital twins, and blockchain. The development of the Metaverse has entered the early stages of explosion.

\textbf{Rise period.} 2021 is considered likely to be the first year of the Metaverse. Roblox went public in March 2021 with the title of "the first stock of the Metaverse", which triggered a heated discussion about the Metaverse. In May 2021, the Google I/O conference announced a 3D video calling technology called Starline. In October 2021, Facebook changed its corporate name to Meta, taking the prefix from the Metaverse, officially starting a comprehensive transformation from a social media platform to the Metaverse ecosystem. In November 2021, numerous companies began to turn their attention to the Metaverse. Microsoft announced at the Ignite conference that it has officially entered the Metaverse field and will launch Mesh\footnote{https://www.microsoft.com/en-us/mesh} for Microsoft Teams software to assist in office work. NVIDIA launched Omniverse\footnote{https://developer.nvidia.com/nvidia-omniverse-platform}, a platform for generating interactive AI avatars. It can help creators build virtual characters in the Metaverse. Disney will take the Metaverse as its future development direction and build a supporting IP image. In January 2022, Microsoft acquired Activision Blizzard, the world's largest game developer and publisher, for \$68.7 billion. So far, the development of the Metaverse has entered a stage of rapid development.

\begin{figure*}[h]
	\centering
	\includegraphics[scale = 0.38]{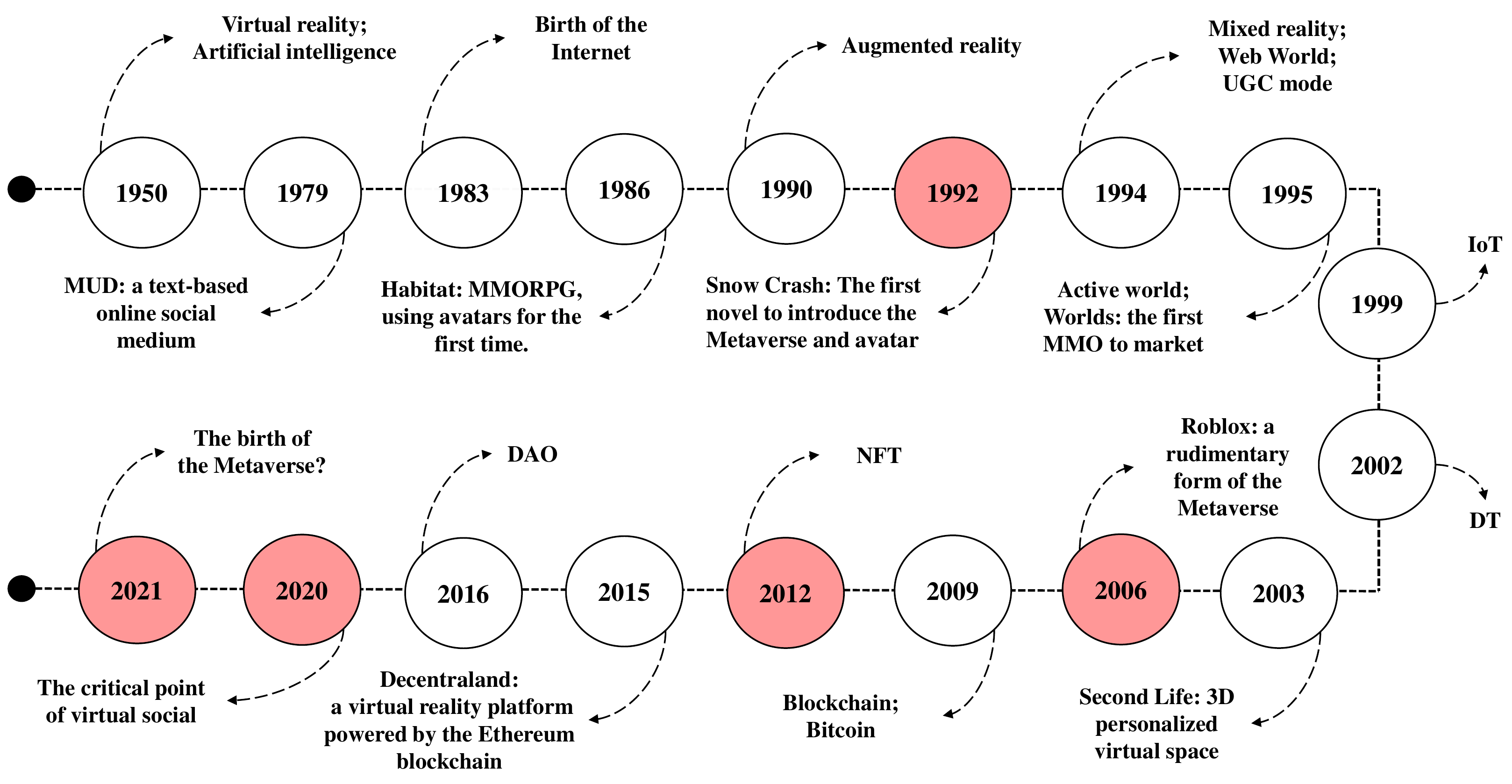}
	\caption{The timeline of the Metaverse}
	\label{fig:timeline of Metaverse}
\end{figure*}

The detailed timeline of the Metaverse is shown in Fig. \ref{fig:timeline of Metaverse}.

%%%%%%%%%%%%%%%%%%   timeline of Meta    %%%%%%%%%%%%%%%%%%
%%%%%%%%%%%%%%%%%%   timeline of Meta   %%%%%%%%%%%%%%%%%%

\subsection{What is the Metaverse?}

The term "Metaverse" is made up of two words: "Meta" and "Verse". Meta means beyond and Verse means universe. It integrates a large number of existing technologies, including 5G, cloud computing, AI, XR, BC, digital currency, IoT, human-computer interaction, etc., and has high requirements for these technologies. As a new type of Internet application, it aims to build a virtual world parallel to the real world with a stable society and economic system, allowing each user to produce content and edit the world. Some call it the new form of the Internet, some call it the digital layer of everyday life, and still others call it a fusion of virtual and physical reality, a persistent virtual space, or a digital twin of the real world. The Metaverse is still an evolving concept, with different players enriching its meaning in their own way.

There are many definitions of the Metaverse with different focuses among the published papers, and several representative definitions are summarized in Table \ref{DefofMeta}.

{\renewcommand\arraystretch{0.8}
	\small
\begin{longtable}{|p{2cm}|p{0.6cm}|p{4.5cm}|p{5.5cm}|}
	\caption{The definition of the Metaverse}
	\label{DefofMeta}\\  \hline  
	\textbf{Source} &\textbf{Year}&\textbf{Definition}&\textbf{Detail}\\  \hline
	\endfirsthead
	\multicolumn{3}{l}{continue\ref{DefofMeta}}\\
	\hline
	\textbf{Source} &\textbf{Year}&\textbf{Definition}&\textbf{Detail}\\  \hline
	\endhead
	\hline
	\endfoot
	Stephenson \cite{stephenson2003snow} & 2003 & In an online world parallel to the real world, people can interact in a three-dimensional virtual space mapped out by the real world through avatars. & A digital space running in parallel with the physical space, with the characteristics of multiple users, persistence, and openness. \\ \hline
	
	Jaynes \textit{et al.} \cite{jaynes2003metaverse} & 2003 & A universal, shared immersive environment that breaks the traditional space-time barriers by tricking the user's visual senses. & Visually immersive, self-configuring and monitoring, interactive, and collaborative. \\ \hline
	
	Wright \textit{et al.} \cite{wright2008augmented} & 2008 & A large 3D online virtual world capable of simultaneously supporting a large number of users for social interaction. & Interaction of real people with virtual environments and avatars with increasing immersion and presence. \\ 
	\hline
	
	Hazan \cite{hazan2010musing} & 2010 & A persistent world where users log in anytime to interact with others in play, commerce, creativity, and exploration. & A persistent world has a dynamic all its own. \\ \hline
	
	Papagiannidis \textit{et al.} \cite{papagiannidis2010staging} & 2010 & A virtual world where economic and social activities can take place, based on the Internet and modeled on the real world. & The Metaverse and its activities interact with other digital spaces and the real world, and its application has important practical implications. \\  \hline
	
	Owens \textit{et al.} \cite{owens2011empirical} & 2011 & A three-dimensional virtual space based on the real world, with immersive, unlimited, interactive characteristics.  & A Metaverse is a shared space that can be dynamically configured. Avatars of the team can interact and cooperate in the Metaverse, personalizing how they choose to work to accomplish tasks. \\	 \hline
	
	Dionisio \textit{et al.} \cite{dionisio20133d} & 2013 & A computer-generated world is a large network of interconnected virtual worlds. & A fully immersive 3D digital environment, reflecting the wholeness of shared online space across all dimensions of presentation, has four features: realism, ubiquity, interoperability, and scalability. \\	 \hline
	
	Rehm \textit{et al.} \cite{rehm2015metaverse} & 2015 & A platform that integrates physics, humans, and technology in the cyber-physics system (CPS). & A tool for cyber-physical evolutionary transformation at every level. It's based on virtual world technologies like extended reality and digital twin.  \\	 \hline
	
	Choi \textit{et al.} \cite{choi2017content} & 2017 & An effective combination of augmented reality and virtual worlds. & Four elements: augmented reality, virtual world, life record, and mirror world. \\	 \hline
	
	Nevelsteen \cite{nevelsteen2018virtual} & 2018 & An interactive human-machine-mediated simulation of an artificial environment where avatars can virtually interact with each other, acting and reacting to things, phenomena, and the environment. & The Metaverse is a real-time shared virtual space in which all user interactions and actions take place. It contains many data spaces that are interconnected. \\	 \hline
	
	Ryskeldiev \textit{et al.} \cite{ryskeldiev2018distributed} & 2018 & An effective combination of augmented reality and virtual worlds. & Reducing the computational costs for mobile mixed reality applications and expanding the available interactive space. \\	 \hline
	
	Falchuk \textit{et al.} \cite{falchuk2018social} & 2018 & A virtual shared space created by the fusion of a virtual reality space and a persistent virtual space. & Implemented by an engine that provides the computational basis for all aspects of the Metaverse (including physics, appearance, communication, synchronization, etc.); sometimes editable, as avatars can affect virtual worlds.   \\ \hline
	
	Huggett \cite{huggett2020virtually} & 2020 & A social virtual world that parallels and, in some ways, replaces the real world. It is the future form of the Internet. & Consists of virtual worlds that combine VR with physical objects, interfaces, and networks. A variant of the Metaverse can be viewed as a virtual world representing a new class of information systems. \\	 \hline
	
	Siyaev \textit{et al.} \cite{siyaev2021towards} & 2021 & People can gather and interact in millions of 3D virtual experiences in mixed reality digital places in the physical world. & Embedding people's lives and creating virtual experiences in the physical world; bringing people into an online digital environment for increasingly shared experiences.  \\	 \hline
	
	Duan \textit{et al.} \cite{duan2021metaverse} & 2021 & A virtual world with human-centered computing. & The Metaverse has the ability to positively impact the real world in terms of accessibility, diversity, equality, and humanity.   \\	 \hline  
\end{longtable}

}
\normalsize	

\subsection{The characteristics of the Metaverse}

\begin{figure*}[h]
	\centering
	\includegraphics[scale = 0.54]{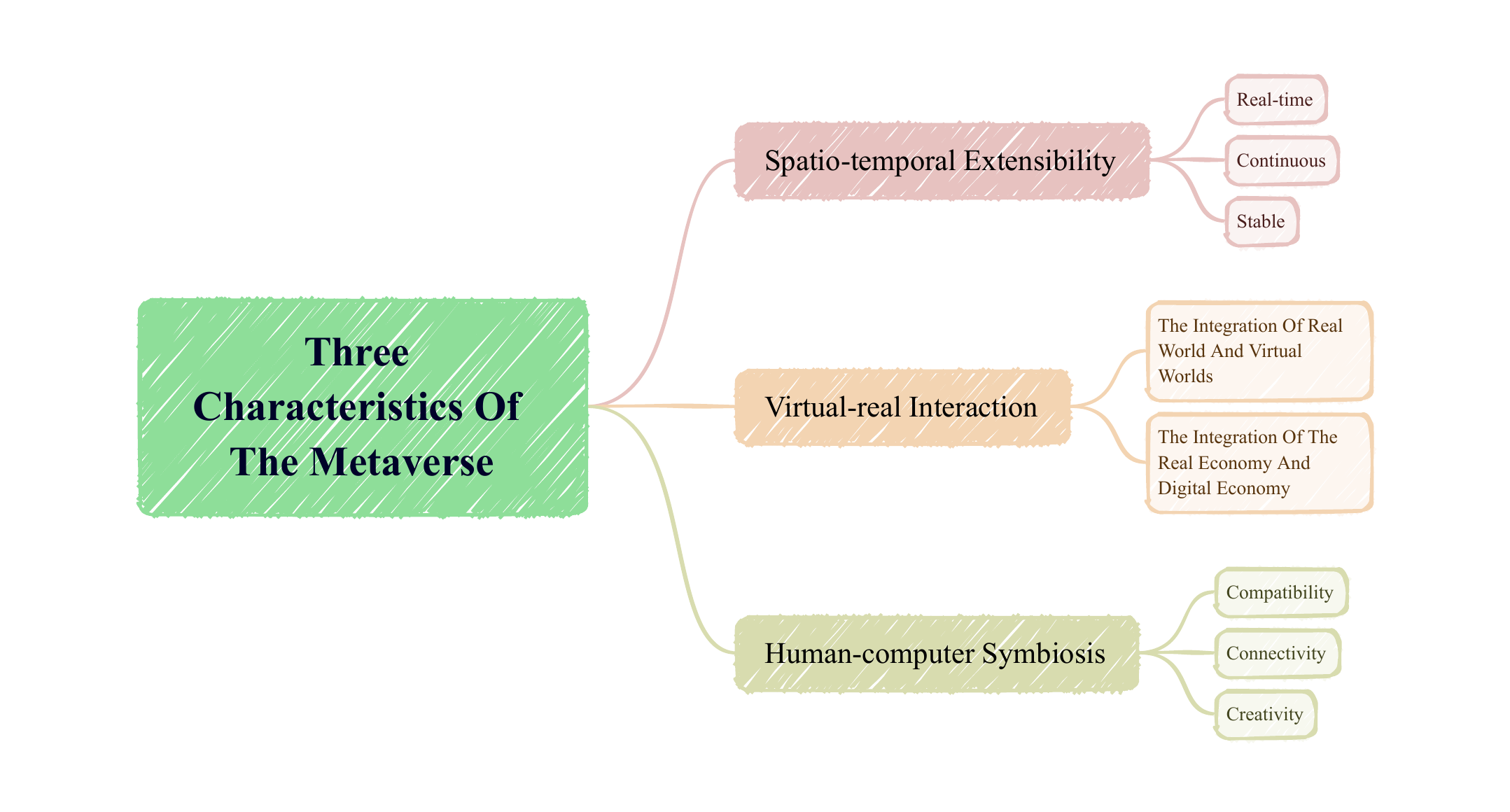}
	\caption{Three characteristics of the Metaverse}
	\label{fig:Characteristics of Metaverse}
\end{figure*}

In this article, the characteristics of the Metaverse are summarized into three: spatio-temporal extensibility, virtual-real interaction and human-Computer symbiosis, as shown in Fig. \ref{fig:Characteristics of Metaverse}. The details are introduced as follows.

\textbf{Spatio-temporal extensibility.} As a virtual space parallel to the real world, the main feature of the Metaverse is its spatio-temporal extensibility. The time and space of the Metaverse are made up of data and algorithms. The most intuitive is the space expandability. At present, the main living space of human beings is the real world. When the Metaverse appeared, the living space of human beings expanded to include the virtual space. Real space and virtual space are interconnected and affect each other. Virtual space is no longer just a simple reproduction of physical space, but a parallel space that can function independently and interact with the real world. Unlike the finite real space, the space in the Metaverse is infinite. It is not only the mapping and expansion of physical space, but also the construction and recording of spiritual space. In the Metaverse, people can reconstruct their memory space or record their dreams. In addition, space in the Metaverse is not constrained by traditional physics, so many things that are difficult to achieve or exist only in the human imagination are possible. The second one is time scalability. Unlike the linear time in nature, the time of the Metaverse is retroactive, and the past, present, and future can be spanned. Simple real-world time simulations, such as year, month, season, etc., can be implemented in the Metaverse. Time can be set by the platform or by the user. Linear time does not work. In terms of time breadth, the information in the Metaverse can be recorded in detail, and users can go back and use it at any time and in any space, making up for the shortcomings that information in the real world is difficult to collect completely.

%%%%%%%%%%%%%%%%%%   Characteristics of Meta    %%%%%%%%%%%%%%%%%%
%%%%%%%%%%%%%%%%%%   Characteristics of Meta   %%%%%%%%%%%%%%%%%%

From a technical perspective, continuity, real-time, and reliability are the necessary features for the Metaverse to meet the spatial-temporal extensibility. Continuity means that the virtual world constructed by the Metaverse can exist forever and will not stop. As a mirror and complement to the real world, the Metaverse will become an important part of human society. Real-time and reliability refer to the ability to maintain real-time synchronization and reliable transmission of information and state with the real world, so as to ensure the timeliness and accuracy of real and virtual interaction. The Metaverse transcends space and time, mapping currency, environment, events, and other information in shared space, and enabling information jumping and virtual time travel.

\textbf{Virtual-real interaction.} Unlike the virtual worlds in most games today, the virtual world and the real world in the Metaverse are interactive, reinforcing and influencing each other. The focus of the Metaverse is the fusion of the virtual and the real. The first is the convergence of the real world and the virtual world. Open-world games can be seen as a precursor to virtual worlds in the Metaverse. An important new feature of the Metaverse compared to games is its persistence, which means that the digital world can continue to exist and evolve in the same way as the physical world. The Metaverse is not just a virtual world, but a strong interaction and deep integration between the digital world and the physical world. Because of this, it can promote the Internet and social economy to a higher level of progress. If the digital world does not add value to the physical world, there is not much room for growth. Digital twin technology can achieve "from the real world to the digital world". At the same time, with the development of XR technology, "from the digital world to the real world" is gradually becoming a reality. The second is the integration of the real economy and the digital economy. According to Matthew Ball, a venture capitalist who has studied the Metaverse for a long time, the Metaverse needs to form a fully mature economy \cite{matthew2020meta}. Virtual and reality need to be interconnected at the economic level to form a complete closed-loop economic system that is highly digital and intelligent. Blockchain-based smart contracts will increase the credibility of Metaverse transactions and greatly reduce transaction costs. At the same time, Metaverse will provide people with low-cost and high-efficiency intelligent financial services, improving the convenience of financial services. Digital content is an important part of the Metaverse, and the Metaverse is a creator-driven world. Digital land, algorithm models, data resources, etc. can form valuable digital assets. Both the related data economy and the creative economy can promote the development of the real economy and build a new digital economic system. The third is the integration of digital life and social life. In the Metaverse, people can try all kinds of ideas that are limited by real life and meet their spiritual needs. People's digital identities will gradually merge with their real identities to form a unified new identity system and build digital credit in the Metaverse. It is entirely possible for digital identities to realize cross-platform interoperability and mutual recognition. Based on digital identity, we can unify identity, assets, and data on the blockchain. All digital assets are based on digital identity management, which can effectively ensure asset security. In addition, blockchain technology can not only allow digital assets to authenticate, circulate, and ensure asset security, but also allow digital economic activities in the Metaverse to accumulate and form a large amount of digital wealth, improving the liquidity of assets.

\textbf{Human-Computer Symbiosis.} The emergence of COVID-19 has created a social distance in people's communication, and the development of digital technology has allowed us to shorten this distance, accelerating the symbiosis between people and digital people and robots. At present, smart media devices have expanded people's abilities to a certain extent, allowing people to have richer experiences and stronger life skills. The increasingly close integration of the human body and technology promotes the evolution of the main body of social activities from humans to the cybernetic organism embedded with humans and machines. With the continuous progress of AI technologies such as computer vision, natural language processing (NLP) \cite{chowdhary2020natural,zhu2022metaaid} and XR, as well as the continuous development of the cross-technology support system formed by integrating hardware, network, and computing, the Metaverse of virtual-real integration and intelligent interaction is becoming a reality. Humans enhance their capabilities through the Internet of Things and robots, which will realize the transformation from the real world to the virtual world. The virtual human, in the form of an avatar, expands human capabilities in virtual space. Humans, robots, and virtual humans coexist in the Metaverse.

There are the following two points about human-computer symbiosis in the Metaverse that need to be considered. First, it is necessary to bring computers into real-time thinking processes. The computer needs to be aligned with the human way of thinking. By computing in real time, the ability of the computer is used to complement the ability of the human, and the final decision on the problem can be obtained. The second is about the specific realization of human-computer symbiosis. With the joint support of new sensing technology and artificial intelligence technology, computers will be able to understand human intentions and the surrounding environment through perception and data processing technology, resulting in more natural and intelligent human-computer interaction. Moreover, the human-computer interaction mode should be multimodal. People can use a keyboard, mouse, and voice to operate the computer, as well as gestures, expressions, and lip language. One of the most important points is to achieve an accurate understanding of user intent through artificial intelligence, machine learning, and other methods. The user's true intention can be inferred by constructing a user-intent understanding framework, modeling user behavior characteristics, and inputting vague signals. In order to solve the uncertainty of human-computer interaction, it is necessary to use artificial intelligence related technologies. With the accelerated integration and innovation of digital technology and the maturity of the cross-technological system, more virtual-real integration and intelligent interactive products for industrial and consumer scenarios will emerge, which will in turn promote the deep integration of the digital economy and the real economy and enrich people's lives.

%%%%%%%%%%%%%%%%%%%%%%%%%%%%% separator %%%%%%%%%%%%%%%%%%%%%%%%%%%%%%%%

\section{Technologies}  \label{sec:Technology}

The construction of the Metaverse requires a variety of new technologies, protocols, companies, and innovations. It does not appear suddenly. In other words, there are no very clear boundaries to define the emergence of the Metaverse. As different technologies, functions, and services evolve and converge, the Metaverse will be gradually built. There are three points that need to be discussed.

\subsection{Framework}
%https://www.matthewball.vc/all/forwardtothemetaverseprimer

\begin{figure*}[h]
	\centering
	\includegraphics[scale = 0.62]{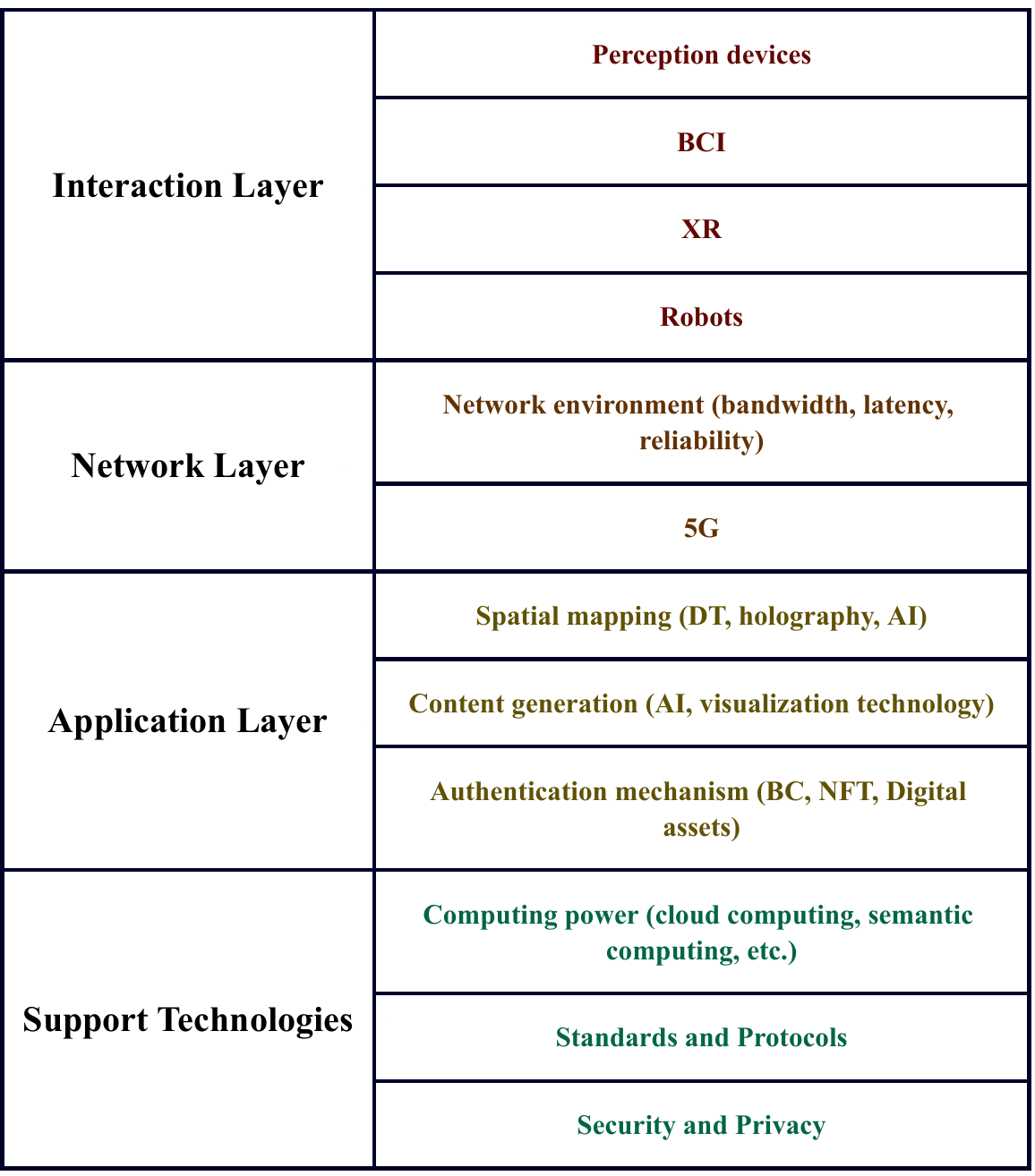}
	\caption{Framework of the Metaverse}
	\label{fig:Framework}
\end{figure*}

The framework of the Metaverse can be summarized into four layers, including the interaction layer, the network layer, the application layer, and the related supporting technologies, as shown in Fig. \ref{fig:Framework}. The main task of the interaction layer is to realize communication between humans and the virtual world and ensure the immersive experience of users. The technologies involved in this layer mainly include perception devices, brain-computer interface (BCI) \cite{saha2021progress,mystakidis2022metaverse}, XR technologies, and robots. The network layer includes the technology and hardware foundation for realizing network transmission, providing a high-bandwidth, low-latency, stable, and secure network environment for the Metaverse. The current major network technologies are 5G and 6G. The application layer is mainly related to the content production and maintenance of the Metaverse, including spatial mapping, content generation, and authentication mechanisms. Spatial mapping is the complete real-time mapping based on the real world, which is related to digital twin (DT), holography, and AI. Content generation is the driving force behind the operation of the Metaverse, which is related to AI and visualization technology. The authentication mechanism includes technologies such as BC and Non-Fungible Token (NFT). It is a guarantee for the stable operation of the Metaverse, providing it with a secure, certifiable digital environment. Furthermore, computing power, standards, and protocols, as well as security measures, are the underlying technologies for the realization of the Metaverse.

%%%%%%%%%%%%%%%%%%   Applications    %%%%%%%%%%%%%%%%%%
%%%%%%%%%%%%%%%%%%   Applications   %%%%%%%%%%%%%%%%%%%

\subsection{Technical support}

\textbf{Network.} The main task of the network is to realize the transmission, reception, and sharing of information between computers. Communication technology and networks are the basis for the Metaverse to form a virtual world that is interconnected with the real world. Its limitations and development affect the design of products and services in the Metaverse, as well as the domains and times in which the Metaverse can be applied. The main indicators used to measure network performance are bandwidth, latency, and reliability \cite{matthew2021net}.

First, bandwidth refers to the maximum number of bits of data flowing through a specific area under conditions such as a given time, which measures the maximum rate at which data can theoretically be transmitted in a network. The complete reconstruction of the real world in the Metaverse is not a one-off, and the data is synchronized and interacts with the real world changes in real time. As the complexity of the virtual simulation increases, the amount of data that needs to be transmitted over the network increases dramatically. The Metaverse needs the network to provide massive connections, high-performance end-to-end transmission delay, in-network computing cache, and flexible access to processing capabilities. Therefore, it is urgent to make changes to the existing network architecture. At present, the representative of communication technology is 5G, so it is an inevitable choice to build a matching infrastructure according to the Metaverse scenario, such as building 5G base stations and other network facilities. Wi-Fi 6, 6G communication, and other technologies are yet to be explored and developed to meet the goal of massive information exchange and timely response in the Metaverse. Secondly, as a super-large virtual twin platform, the Metaverse must ensure low latency of the network, which can guarantee the social experience and expectations of users. Current social applications do not have a high demand for network latency, such as synchronized video calls. However, the social nature of the Metaverse and the need to cater to a wide variety of scenarios means it requires low latency. The social activities and remote technical operations of the Metaverse are very sensitive to minor errors and synchronization issues. The goal of realizing XR, tactile synergy, holographic perception communication, and other technologies in the Metaverse is to achieve 1 ms end-to-end latency. In fact, latency is the hardest of all the network properties to fix. When using edge computing, the construction of edge computing nodes can shorten the distance of information flow transmission, thereby reducing the transmission delay of the Metaverse network. Finally, the reliability of the network is the basis for the stable operation of the Metaverse. Whether it is telemedicine, autonomous driving, virtual labor, or virtual education, it is very necessary to use a reasonable network structure and fault-tolerant design to ensure the reliability of services.

\textbf{Compute.} As the carrier of the digital world, the Metaverse needs powerful computing power to satisfy various functions and services, such as physical computing, rendering, data synchronization, etc. Computing power will determine the upper limit of the size of the Metaverse. As a mirror world that operates all the time, the Metaverse has enormous continuous computing demands. At the same time, the modeling of large-scale and highly complex digital twin spaces and digital humans requires the collaborative creation of many designers, and the interaction between the real world and the digital world requires real-time, high-definition 3D rendering computing power and low-latency network data transmission. It raises the processing requirements of cloud collaboration as well as the enormous amounts of graphics and image computing required. The application of the Metaverse will involve multiple types of physical simulations such as power, heat, and fluids, which requires the use of high-precision numerical calculations to support physical simulation and scientific visualization. Meanwhile, the Metaverse will also involve AI application scenarios such as human-computer interaction. AI-driven digital humans often need to combine AI algorithms such as speech recognition and NLP to achieve interactive capabilities. These models require powerful AI computing power to support their training needs. The construction of a highly realistic digital world and the realization of real-time interaction between hundreds of millions of users is faced with many challenges, including large scene scale, high scene complexity, real-time rendering, simulation, interaction, and many other challenges, and the core power supporting the construction and operation of the Metaverse - computing power imposes higher requirements. This requirement is not only a high-performance, low-latency, and easily extensible hardware platform, but also an end-to-end, ecologically-rich, and easy-to-use software stack.

The development of hardware computing capabilities and edge cloud computing capabilities will further enhance users' low-latency and high-fidelity experiences. First, the improvement of hardware computing power, especially GPU computing power, can further enhance the display effect of the Metaverse and cloud games, making it possible to model more realistic scenes and items. Secondly, through edge cloud computing, the performance requirements of terminal equipment can be reduced. With edge computing, networked devices process data at edge data centers or locally, without transferring data back to a central server. This will reduce the load on the cloud and increase processing speed significantly, providing faster responses to users. Meanwhile, the security of the network has improved. Edge computing distributes data processing among different data centers and devices. Hackers cannot affect the entire network by attacking just one device. In addition, cloud computing systems that dynamically allocate computing power will be part of the infrastructure of the Metaverse, including cloud storage and cloud rendering.

\textbf{Virtual and real interactive interface.} In real life, the way users access the Metaverse should be easy to operate, convenient, and adaptable. At present, there are mainly XR, BCI, robots, and so on to connect reality and the Metaverse. XR includes VR, augmented reality (AR) \cite{zhan2020augmented,siriwardhana2021survey}, and MR. Through information technology, VR is a technology that presents a virtual environment similar to or different from the real world to users. AR superimposes a layer of virtual information on the basis of retaining the real world, allowing users to obtain data and information that has been analyzed and processed by computers in real time, which can help users in their work and decision-making. MR refers to a new visual environment created by merging the real world and the virtual world, and is a merger of VR and AR. XR technology realizes the input and output of information in the Metaverse by capturing user actions and using the user's vision, hearing, and touch. These XR technologies belong to immersive media, which can present digital content in the Metaverse to users from a first-person perspective, and are the supporting technologies for realizing virtual-real interaction. Currently, XR technologies rely on screens and traditional control systems, some of which also use touch and smell. However, to achieve stable long-term virtual worlds, brain-computer interfaces need to be explored and utilized. The goal of BCI is to completely replace screens and physical hardware, leveraging the user's mind to drive applications. The application of BCI is becoming the focus of competition among technology giants, and it is currently mainly used in the medical field. In the future, the development of BCI enables human thoughts to be tracked, recorded, and shared. Robots, as physical simulations, can expand human capabilities in the real world and become another channel to connect the Metaverse. Synthetic data generated from the virtual world may guide robots to solve problems and replace humans in high-risk jobs.

\textbf{Content production.} An important part of the services provided by the Metaverse is its content, including content generation, presentation, and review. The first is content generation. Building a Metaverse that is highly consistent with the real world or even surpasses the real world requires a lot of data simulation and powerful computing power to create a virtual world, the key core of which is the digital twin (DT). The DT is the cornerstone of the Metaverse. A digital twin is a dynamic twin that creates a real thing in a virtual space \cite{tao2018digital,rasheed2020digital,fuller2020digital,liu2021review}. With the help of sensors, the operating state of the ontology and external environmental data can be mapped to the twin in real time. It is a virtual representation of an object or system that spans its lifecycle and uses techniques such as machine learning to aid decision-making, originally used in industrial manufacturing \cite{ibm2021dt}. Virtual models accurately reflect real objects or systems through sensors that actively forward data related to their function and environment to their digital twins in real-time. Any change in a physical object or system results in a change in the digital representation and vice versa. The system uses methods of data analysis to predict when the machine will fail and repair it before it occurs. The Metaverse needs the DT to build a realistic environment with extremely rich details, creating an immersive experience. The maturity of the DT determines the integrity that the Metaverse can support in its virtual-real mapping and interaction.

Simulating and synchronizing the real world is only the first step in building the Metaverse, and the massive amount of content available to users without repetition is a key component of the Metaverse service. The intelligence level of digital content creation is constantly improving with the development of AI. From the original professional-generated content (PGC), to user generated content (UGC) \cite{ondrejka2004escaping,lastowka2007user,naab2017studies} and artificial intelligence-generated content (AIGC), to the fully automated AIGC in the future, while the cost of content creation is decreasing, the requirements for intelligence are gradually increasing \cite{ondrejka2004escaping}. The content created and generated by AI is an important productive force for future industries. Low-code and automated content will become mainstream, and it will greatly reduce the threshold for creation and the cost of creation. The creative market will shift from niche to mass.

Secondly, it is worth thinking about how to present the content of the Metaverse to users reasonably. DT provides a platform for content presentation. The Metaverse uses a digital copy of the real world built by DT as the entry point to provide users with a fully connected 3D experience. At the same time, XR technology and various sensors are used to optimize the physical world and enhance the user's sense of immersion in the real world. Digital avatars powered by AI present the contents of the Metaverse to the user in an organized way. AI is used to process large amounts of unstructured data and improve the efficiency of computing in complex environments. By expanding the knowledge graph, AI can participate in the whole process of decision-making, planning, execution and adjustment, further improving the level of automation, so as to comprehensively improve the decision-making intelligence of complex systems. Finally, content review is also necessary. AI technology is used to review the massive contents that cannot be manually completed in the Metaverse to ensure the safety and legality of the Metaverse.

\textbf{Authentication mechanism.} Unlike human society, the Metaverse is completely digital. Therefore, a new authentication mechanism is needed to build a unique credit system in the Metaverse and solve the centralization problem inherited from the traditional Internet. On the one hand, a stable economic system can encourage users to generate content independently, so as to ensure that there is rich content in the Metaverse for users to use and consume. In the economic system, the credit system is indispensable. On the other hand, the digital world is characterized by centralization, easy replication, and easy dissemination. Centralized systems have great risks in the creation and operation of virtual worlds, such as hacking, malware, and centralized decision-making. The authentication mechanism of the real world is ineffective in the digital world, and it is difficult to protect the rights of creators and users. BC minimizes these risks and makes it possible to build a stable virtual world. The distributed ledger of the BC realizes that each node stores complete data. This data cannot be tampered with, cannot be deleted, and can be traced back in time. The BC relies on the consensus mechanism and asymmetric encryption to ensure the consistency of storage and the security of user data. Based on this trusted and immutable data, smart contracts can effectively regulate the economic, legal, social, and other relationships between participants within the Metaverse. Non-Fungible Token (NFT) \cite{wang2021non,vidal2022new} is a digital asset right certificate based on BC. After the digital asset is minted into NFT, it will be permanently stored on the BC, which is unique and immutable. In general, building a true Metaverse requires addressing the issues of assets and identities in the system. Everyone's assets, identities, and other data are recorded in a decentralized manner through the BC technology. At the same time, each virtual asset records and confirms the ownership through the NFT contract and will not lose value due to being copied or counterfeited. The virtual currency based on the decentralized network makes the attribution, circulation, and realization of the value and the authentication of the virtual identity possible in the Metaverse.

\section{Applications}
\label{sec:Applications}

\begin{figure*}[h]
	\centering
	\includegraphics[scale = 0.56]{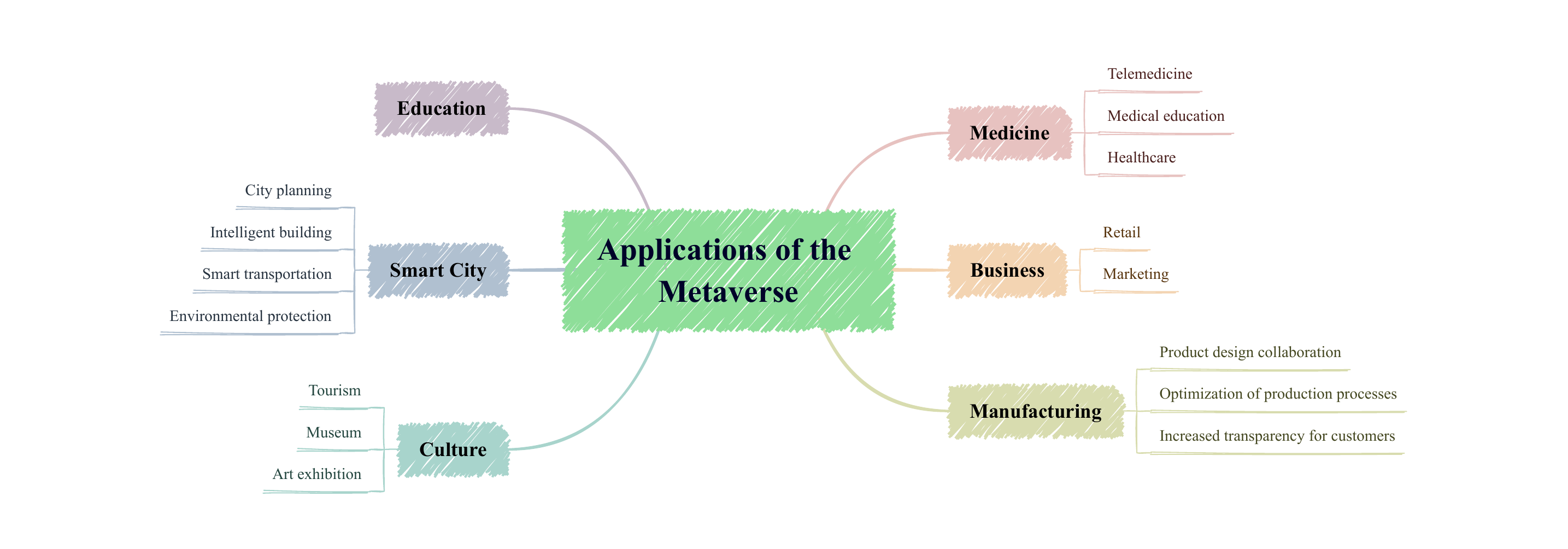}
	\caption{Applications of the Metaverse}
	\label{fig:Applications}
\end{figure*}

This article mainly summarizes the application status of the following six fields, as shown in Fig. \ref{fig:Applications}. Metaverse applications in other fields are also listed in Table \ref{tab:AppofIoB}.

%%%%%%%%%%%%%%%%%%   Applications    %%%%%%%%%%%%%%%%%%
%%%%%%%%%%%%%%%%%%   Applications   %%%%%%%%%%%%%%%%%%%
	
\subsection{Education}
	
Before the COVID-19 pandemic, online education was growing steadily every year. Effective teaching in a virtual environment has always been a research hotspot \cite{ayiter2008integrative, collins2008looking, diaz2020virtual, suh2022utilizing}. As the COVID-19 pandemic spreads across the globe, the school lives of many children, especially those in vulnerable situations, have been disrupted. According to UNICEF, in April 2020, offline teaching has been affected in most countries, with more than 91\% of students worldwide unable to take offline courses \cite{jason2020keeping}. The emergence of the pandemic has limited the movement of people in physical spaces and dramatically accelerated the growth of digital content. Online teaching can solve the inconvenience caused by geographical restrictions, but traditional methods often fail to achieve good results. For example, in a game environment, students will be distracted. The communication between teachers and students through electronic screens and voice does not have the same convenience and effectiveness as in the real world \cite{maharg2007simulations, kanematsu2009practice}. With the development of various digital technologies such as computing power and networks, the use of immersive virtual environments in the field of education has gradually become a reality. Through XR, cloud computing, and other technologies, teachers, coaches, and experts are brought together with students in the same virtual learning space \cite{diaz2020virtual, mystakidis2022metaverse, hirsh2022whole}. Real interactions between teachers and students are combined with digital twins of avatars or devices to visualize virtual teaching. While breaking physical distance restrictions, teaching efficiency has also been greatly improved.
	
On the basis of restoring the traditional classroom, the application of the Metaverse can achieve many things that are difficult or impossible to achieve under realistic conditions. In the Metaverse, for example, students can perform experiments involving dangerous chemical reactions on their own, which is also conducive to the development of safety education \cite{kanematsu2014virtual}. Medical experiments often require a large number of conditions and materials that are difficult to meet, but students need to repeat the experiments many times to gain proficiency. A virtual experiment environment that meets accuracy standards can meet high-standard teaching requirements while reducing educational costs \cite{almarzouqi2022prediction, locurcio2022dental}. In addition, the question of education for children with disabilities may be answered in the Metaverse. Remote teaching of sign language using avatars that simulate human gestures based on AI systems could make the learning process for hearing-impaired children easier and more fun \cite{batnasan2022arsl}. It should be noted that when designing education-related virtual environments, students' learning laws and principles should be followed, rather than just pursuing interesting things to attract students' attention \cite{hirsh2022whole}. Too much interaction may cause students to be distracted, while ignoring knowledge acquisition and ability training. Currently, there is still a lot to build and explore in the educational Metaverse. The representation of the Metaverse and its accessibility are the top priorities.
	
\subsection{Medicine} 
	
How to keep healthy is a very important topic in people's lives, so people have been pursuing continuous innovation and development in the medical field. From MRI scans and X-rays to robotic surgery and VR, the healthcare industry is undergoing a massive digital transformation \cite{YU2012Building, Claudio2014Overview, Pensieri2016Virtual, Alonso2021Telemedicine}. The realization of telemedicine, medical education, and healthcare in a 3D virtual environment can overcome the limitations brought by physical factors and bring a new research perspective for medical research.
	
\textbf{Telemedicine.} The term "metaverse" became a recurring hot topic in 2021, and its usage has exploded. As a combination of various digital technologies, the Metaverse is well positioned to interconnect physical and virtual objects, including spaces and objects used in the delivery of real-time medical services. In the virtual world, the use of avatar-based doctor-patient consultation breaks through geographical constraints and can also directly integrate existing telemedicine services in the Metaverse. Yang \textit{et al.} \cite{yang2022expert} proposed a definition of the medical Metaverse, the Medical Internet of Things (MIoT) using AR and VR technologies. 2022 is proposed to be named the year of the Metaverse in Medicine. They believe that the three basic functions of MIoT are comprehensive perception, reliable transmission, and intelligent processing. Skalidis \textit{et al.} \cite{skalidis2022cardioverse} proposed CardioVerse (a term used to describe the concept of the Metaverse adopted by cardiovascular medicine), emphasizing the feasibility and importance of the Metaverse in assisting cardiovascular therapy. The main advantage of the Metaverse is that it is highly realistic. Through high-resolution micro-CT scans and 3D modeling technology, doctors can conduct meetings within virtual organs. This approach deepens researchers' understanding of human organs from a new perspective and improves the efficiency of disease treatment \cite{werner2022use}. Using Microsoft HoloLens technology as a surgical aid, surgeons can more accurately detect, diagnose, and treat patients. 
	
\textbf{Medical education.} The use of Metaverse to conduct simulated training for students breaks through the limitation of one-way knowledge dissemination in traditional teaching methods. The innovation and use of MR in education will greatly influence teaching practice and achieve better teaching results \cite{almarzouqi2022prediction}. With the help of sensors and other equipment to simulate the real surgical process, students can better learn knowledge and skills in the process of practice. For example, the use of haptic gloves can allow dental students to feel the immediate feedback of virtual objects, such as needle tip insertion, as they practice suturing, thereby gradually improving the students' technical ability \cite{locurcio2022dental}. More importantly, the immersive experience and interoperability of the Metaverse can open up new avenues for medical training. Through the combined use of MR and AI, students are able to enter a virtual human body to obtain a complete view of human organs or simulate actual surgery \cite{skalidis2022cardioverse}. Different types of virtual worlds have been introduced into medical education, including AR, life logs, and VR \cite{huh2022application}. While surgery requires not only adapting to the virtual world, but also the use of flexible grasping instruments. This requires the use of appropriate hardware or tracking technology, so the Metaverse technologies must be more flexible and adaptable to human operation.
	
\textbf{Healthcare}. Firstly, by providing patients with a relaxed and pleasant virtual environment during their treatment or rehabilitation, the pain and anxiety of the patients can be relieved, which is conducive to the early recovery of the patients. Second, blockchain is a key part of the healthcare sector. It democratically controls decentralized communities through smart contracts, as well as records ownership of all data in the digital world \cite{yang2022fusing}. Currently, health data is often transferred between many companies in an inefficient and opaque manner. Since health records are usually kept on a central computer, the data is vulnerable to theft. Blockchain can provide solutions for the management and security of health data. Converting patient data into NFTs could fundamentally improve the security and reliability of sharing and managing electronic health records \cite{wiederhold2022metaverse}. Additionally, in the virtual world, a patient's own digital twin is built for the patient to serve as a personal "test dummy" to predict how the patient will recover from surgery or personalize the treatment plan \cite{next2022meta}.
	
\subsection{Smart City} 
	
Smart cities use various software, communication networks, and the Internet of Things to collect various data in the city, and analyze this data to obtain intelligent solutions to improve public life. Many countries around the world have already begun to build smart cities, aiming to improve traffic conditions, save energy, improve urban safety indexes, and provide optimal planning solutions for urban construction. Smart city technology can not only improve efficiency in manufacturing, urban agriculture, energy use, etc., but also connect various services to provide joint solutions for citizens. In the Metaverse, the use of DT to build digital cities and real-time simulation of real urban environments can promote economic development, realize effective management of human resources, and improve the ecological environment, thereby improving the overall quality of life for residents \cite{hudson2007digital, aurigi2022smart}.
	
\textbf{City planning.} Traditional urban planning methods require on-the-spot investigations. At the same time, data collection and program design will consume a lot of manpower and material resources. Based on DT and XR technology, the city mirror constructed by Metaverse through data can reflect the real-time urban situation in the mirror through IoT technology, and calculate and simulate complex scenes through data processing \cite{shahat2021city,deng2021systematic}. This method will completely reconstruct the logic of urban construction and operation, optimize the layout of urban design, and efficiently guide urban construction in terms of simulated planning and optimized governance. Due to its real-time nature, the Metaverse can provide urban planners with a comprehensive and dynamic simulator, allowing them to obtain real-time status data and environmental data for each house in a virtual city. Appropriate urban planning schemes are obtained through artificial intelligence algorithms to improve decision-making efficiency.
	
\textbf{Intelligent building.} Building Information Modeling (BIM) technology \cite{volk2014building,tang2019review} is an existing collaborative approach in the intelligent building field, which is often used in building design and construction stages. It allows the digital construction of an accurate virtual model of a building that exists throughout the life of a building project. Collaborators plan, design, and build structures or buildings in a 3D model for multidisciplinary information storage, sharing, exchange, and management \cite{eastman2011bim}. By incorporating BIM technology into the Metaverse, practitioners will gain a more holistic perspective and improve design and construction efficiency. The Metaverse can also support the operation and maintenance of intelligent buildings. How to ensure the rationality of the internal layout of the building is an important and complex issue. For example, in emergencies, exit signs can help residents effectively escape the building, and their location selection is very important \cite{fu2020bim}. Static exit signs are generally determined by designers, but with the increasing structural complexity of buildings, it is very difficult to manually determine the location of exit signs. Moreover, static signs cannot solve the problem of possible blockage of evacuation routes. The dynamic exit sign was brought up. It can monitor the environment of buildings in real time and provide residents with optimal evacuation paths based on intelligent algorithms \cite{galea2017evaluating}. In the Metaverse, the routing algorithm is further optimized by simulating evacuation situations in advance to maximize the escape rate of residents in the event of an emergency. The management and maintenance of buildings can also be optimized in the Metaverse. Existing building maintenance solutions solve existing problems through artificial offline or online repair methods, which cannot achieve real-time monitoring of buildings, and lack predictability and efficiency. Buildings in the Metaverse can provide real-time space management or structural health monitoring and feedback to determine when repairs are needed. Utilize real-world sensors to monitor infrastructure issues, such as leaking water pipes, to provide timely feedback to management. Through comprehensive monitoring and analysis of data in the building, early warning of safety risks is realized. Furthermore, buildings are one of the main energy consumers in cities \cite{omar2018intelligent}. How to reduce energy consumption while meeting the comfort conditions of buildings needs to be studied in depth. Through the combination of AI algorithms, ML, and IoT technology, the real buildings are modeled, and the mapping is formed in the virtual world. In this way, it is more convenient to dynamically adjust the building's comfort according to human behavior and improve energy efficiency \cite{ruohomaki2018smart,merabet2021intelligent}.
	
\textbf{Smart transportation.} In the Metaverse, using technologies such as DT, ML, and IoT, traffic data is collected in real time and input into the established traffic model system, realizing a digital virtual mapping of the traffic system. Big data analysis, artificial intelligence, and traffic simulation technology can evaluate existing schemes and obtain traffic optimization schemes. The transportation system in the Metaverse must be a digital twin \cite{deng2021systematic}, which can comprehensively cover traffic management, supervision, planning and design, and transportation services. While managing virtual traffic in the Metaverse, the system can enhance a deeper understanding of the real physical world and make more accurate predictions, enabling true artificial intelligence in traffic management. For example, in the construction of transportation infrastructure, big data analysis is widely used in the efficient design and planning of intelligent transportation to form a reasonable urban transportation layout \cite{jan2019designing}. In traffic management, smart transportation can monitor traffic flow and optimize traffic lights to reduce congestion, while providing drivers and passengers with information available in real-time to reduce accidents and improve road safety. To a large extent, in the IoT-based intelligent transportation environment, proper preprocessing, real-time analysis, and communication models can realize friendly communication in road traffic and avoid road accidents to a large extent \cite{babar2019real}. With more and more vehicles in the city, intelligent parking has become very important. At present, there is related research on parking management using the advantages of the IoT \cite{saarika2017smart}. Drivers can easily find parking spaces, and digital payments are also supported. Parking signs intelligently give useful information about weather and distance. Additionally, traffic safety awareness and education, as well as traffic skills training, will gain new meanings in the Metaverse. Using the AR model, pilots can easily realize various airports and routes, simulate flight scenarios in various weather and time zones, and save a lot of training costs. Traffic safety games help children experience the right way to travel safely in traffic.

\subsection{Business}
	
\textbf{Retail.} There are many Metaverse-related business opportunities in retail. First, Metaverse provides technology and a platform for the convergence of online and offline sales \cite{rehm2015metaverse}. Businesses in the retail industry can leverage digital technologies to create immersive shopping experiences for consumers. Combining real-time commerce with Metaverse using digital twin can overcome the limitations of existing online shopping \cite{jeong2022innovative}. Users can visit stores in the virtual world as digital avatars and make purchases as if they were in the real world. Unlike visiting online shopping sites, users can not only view products, but also try products through digital avatars in a virtual world. At the same time, AR technology allows consumers to fully understand the quality and style of products before purchasing, which can effectively reduce the return rate and provide convenience for both merchants and consumers. With the development of these emerging technologies, e-commerce has shifted from a "click-to-buy" approach to an "experience-to-buy" approach. This will allow Metaverse customers to explore the store, view product displays and make purchases as usual at their home. It's a unique blend of the immersive nature of offline commerce and the convenience of online shopping. On top of that, the combination of Metaverse and commerce also provides a smarter solution for introducing new and complex products. For example, retailers can get timely and helpful feedback on product improvements by letting customers virtually try new products. Metaverse can provide users with a completely different experience and improve design services, increase customer engagement, and enhance communication \cite{gadalla2013metaverse}. Using natural language processing, data-driven decision-making, and real-time IoT data, companies can make optimal business decisions such as design options to ensure brand awareness and customer loyalty \cite{watson2022virtual}.

\textbf{Marketing.} There is great potential for advertising and marketing using the Metaverse in the future. The important factors driving the development of advertising in the virtual world are accessibility and diversity. Metaverse can break the limitations of existing advertising formats by letting users create their own experiences. Businesses can sell virtual counterparts of almost any product they sell in the real world and use the Metaverse for advertising. Using AR technology, users can experience new products anytime in the virtual world \cite{hollensen2022metaverse}. For example, consumers are more willing to try new clothing lines than to see online ads featuring celebrities wearing them. Many brands have recreated their services in the virtual environment of online games. Interestingly, the brand itself is integrated into the gaming environment rather than disrupting the gamer experience. Two of the most popular brands plugged into online games right now include Fortnite and Animal Crossing. In addition, AI algorithms can explore the habits and attitudes of consumers by analyzing the massive data in the Metaverse, providing enterprises with optimal marketing strategies and reducing marketing costs \cite{park2022metaverse}.
	
\subsection{Culture}
	
\textbf{Tourism.} As a powerful tool to virtualize and digitize the real world, Virtual reality technology and the immersive experiences contained in the Metaverse have already been used in tourism, resulting in economic and social effects \cite{pencarelli2020digital, akhtar2021post}. The integration of the Metaverse and tourism can enhance the tourism experience and change the traditional tourism approaches, such as the virtualization of tourist scenery and scenes, as well as the interactive participation and immersive experience of tourists. With the continuous development of 5G, artificial intelligence, AR, VR, and other technologies, the digital transformation of the cultural and tourism industries is the general trend. The COVID-19 pandemic has also accelerated the digital transformation of tourism. Cloud tourism and cloud exhibitions have become a trend. With the help of AR, VR, and other technologies, tourism is no longer single. Immersive and delay-free social interaction is the strength of the Metaverse, making interactions more real, more engaging, and more efficient. Travel based on the Metaverse will break through the limitations of time and space of traditional tourism and achieve a more immersive experience \cite{buhalis2022mixed}. For example, scenes, characters, and plots that can only be described by literature, film, and television, or games, have a new way of approaching the realistic experience in the Metaverse. People will no longer consider the influence of weather, traffic, and distance, and they can go to different places without leaving home, avoiding crowded crowds and long-distance fatigue. Using MR to build fully immersive and 1:1 scale interactive virtual reality models of ancient buildings, Metaverse not only provides visitors with an enhanced experience of cultural heritage sites, but also helps architecture schools build architectural models for teaching purposes \cite{gaafar2021metaverse}. In addition, Metaverse will likely reshape the model of the tourism industry. The Metaverse makes possible the virtual reconstruction of scenic spots and the digital reconstruction of cultural creations. On the one hand, it solves the problem that some tourist attractions are not renewable, and on the other hand, it makes cultural relics have the value of digital assets through reconstruction. This will promote the digital transformation of the cultural tourism industry and form a digital cultural tourism ecosystem. The Metaverse will go beyond the previous experience in terms of virtual reality, immersion, and connection, redefining and changing the travel habits and consumption habits of consumers.
	
\textbf{Museum.} Digital museums break away from the traditional gallery format and offer visitors novel social and cultural experiences in a virtual space \cite{hazan2010musing}. Many organizations and companies are beginning to explore the feasibility of digitizing museums. From collections and exhibitions to social interactions and user experiences, the Metaverse offers great solutions, including online exhibitions, digital collections, immersive experiences, and gamified participation. Since the beginning of the COVID-19 pandemic, attendance at the world's leading museums has declined significantly, increasing the urgency for the cultural sector to diversify its revenue streams. At the same time, consumer demand for different types of visitor experiences is gradually rising \cite{ko2021study}. To survive in the long term, museums must be able to meet the needs of visitors for digital content as well as at-home experiences. Gamified exhibitions can appeal to a broader and younger audience than traditional exhibitions. Many other museums have been experimenting with video games, such as Minecraft\footnote{https://www.minecraft.net/}. Museums around the world are already elevating their digital and virtual content game.
	
\textbf{Art and exhibition.} The Metaverse presents unprecedented opportunities for artists by providing technologies that fully integrate the physical environment with digital creativity \cite{lee2021creators}. The Metaverse's virtual world format facilitates the creation of novel creative genres, such as immersive art, robotic art, and other user-centered approaches. To deal with unexpected situations, the company can adopt a strategy of holding art exhibitions both online and offline. VR galleries, AR maps, and online audio commentary enrich the user's sense of the visit. AR art maps will help people find exhibitors in the city more efficiently and smoothly. Zaha Hadid Architects presents NFTism, a virtual art gallery. The gallery combines MMOG and interactive services. Visitors can access this immersive and interactive 3D world through a wide variety of devices. The application example combines cyberspace, interactive experiences, and economic infrastructure. However, the combination of art and Metaverse requires both technical aspects and issues such as digital privacy and ownership identification of digital artworks, which need to be addressed at the same time.

% Zaha Hadid Architects presents NFTism, a virtual art gallery \cite{dima2021mesh}

\subsection{Manufacturing}
	
Since the beginning of the industrial revolution, the effective combination of systems and machines has been able to increase productivity, reduce product costs, and find new ways to organize work. The digital transformation continues this trend by providing a better understanding of factory performance through digital operations. However, for the most part, the physical world still takes precedence over the digital world. The rise of the Metaverse allows managers to have primary access to the digital space. In the case of manufacturing, one is able to transform this digital space into the physical world, not just enhance it \cite{siyaev2021towards,cai2022compute}.
	
\textbf{Product design collaboration.} In a Metaverse where designers can collaborate on product design from around the world, VR and AR will improve and simplify the product design process in virtual worlds. As with any new technology that allows easy access to user-generated content, more content in specific types and business areas is expected. With more specific measurements and advanced software, the barriers to entry for designing low-cost, easy-to-build products will be significantly lower, and the number of product designs will increase. The Metaverse is an open and interactive public space. As a result, engineers can present the designed product in a virtual environment, making adjustments and modifications based on feedback from the manufacturer, which will shorten the life cycle of the project. Designers and engineers can build corresponding DT models for products and simulate product operation to determine if the product design makes sense and is efficient, without having to perform time-consuming and expensive physical tests \cite{kritzinger2018digital,zhang2020manufacturing}. With more detailed and physics-based design, production errors will be greatly reduced and will reduce the risk of quality control \cite{cai2022compute}.
	
\textbf{Optimization of production processes.} The Metaverse brings together data from different sources to present production processes in a real-time visualization, providing leaders with a fresh perspective on business processes. Engineers can use these virtual spaces to monitor performance, identify problems, and even fix them at a very granular level. This has obvious advantages for hard-to-reach locations, such as mines and oil fields. Engineers can work remotely, advise field technicians in many of these locations, and perform certain tasks themselves \cite{lee2011self}. Meanwhile, Metaverse simulations will allow manufacturers to test thousands of potential plant scenarios and be able to predict the results of scaling up or down. These simulations can identify ways to automate and optimize facilities. Manufacturers use XR to identify equipment problems in a timely manner, enabling effective quality control and maintenance checks \cite{siyaev2021neuro}. This will help them reduce return rates for defective products and lower maintenance costs. NVIDIA has introduced Omniverse, a VR-based collaboration tool for design team collaboration and 3D simulation with a multi-GPU real-time development platform. It is now successfully used in industrial applications.
	
\textbf{Increased transparency for customers.} The building, distribution, and sale of products through 3D representations enables increased visibility into supply chain processes for customers in the Metaverse. Customers can track their orders throughout the entire cycle, from raw materials to finished goods. Greater supply chain transparency will also provide customers with greater visibility into delivery times and shipping costs. Customers will be able to know the exact delivery time of their goods and possible shipping delays, allowing them to prepare for unexpected situations.

\section{Security and Privacy}
\label{sec:Security and Privacy}
	
The Metaverse is an open, shared, and persistent digital environment that includes products, entertainment, workspaces, commerce, and many other aspects of the real world. As more and more human activities take place in virtual spaces, user information, sensitive data, and immersive experiences will require stronger encryption and protection from attacks. Therefore, issues and solutions related to security and privacy will become more important. Like other new technologies, the Metaverse presents considerable difficulties and challenges in terms of security and privacy \cite{wang2022survey,zhao2022metaverse,di2021metaverse}. Because the Metaverse is in its infancy, this provides an opportunity to develop technology with security at the forefront of the design process. The introduction of security protection technologies such as more efficient blockchain networks can provide solutions to security threats, and the development of blockchain and crypto space will provide abundant opportunities for the development of Metaverse \cite{mohanta2019blockchain}. Of all the developments in the Metaverse, security issues play a crucial role in defining the future direction of Metaverse development, so it is reasonable to consider potential security issues related to the Metaverse.
	
\small
\begin{table}[h]
	\caption{The applications of the Metaverse}
	\label{tab:AppofIoB}
	\scalebox{0.9}{
		\begin{tabular}{|l|p{2.5cm}|p{11.5cm}|}
			\toprule   
			\textbf{No.} &\textbf{Application}&\textbf{Description}\\
			\midrule 
			1 & Telecommuting & In the Metaverse, people can create personalized work spaces or virtual offices. More flexible work environments can foster collaboration, regardless of geography. Digital assets of representatives (such as avatars) and objects can be incorporated into 3D spaces as needed, providing a new dimension to online meetings \cite{cai2022compute}. Meanwhile, Metaverse-based telecommuting has contributed to the sustainable development of megacities, alleviating their population pressure \cite{choi2022working}. For example, Microsoft Mesh provides a virtual collaboration platform for telecommuting. Employees work and communicate through digital avatars in virtual spaces integrated with cross-enterprise platforms and information systems to accelerate problem-solving and decision-making. \\ \hline
			
			2 & Real estate & Real estate companies will use the Metaverse to make it easier for buyers to view and experience real estate without having to visit the site \cite{nalbant2021computer}. Interior designers and architects will also be able to simulate the reality of their designs and avoid mistakes. During interior decoration, residents can simulate interior design elements such as furniture placement in advance and consider their rationality more comprehensively. Decentraland\tablefootnote{https://decentraland.org/} has implemented the virtual real estate concept in Metaverse by combining virtual reality and blockchain technology \cite{jeon2022blockchain}. Users are free to place buildings on the land they have purchased and can profit from advertising or exhibitions.  \\ \hline
			
			3 & Social good & The envisioned Metaverse will help people build a fair and sustainable society, weakening the effects of race, gender, disability, and property on equality \cite{duan2021metaverse}. As an autonomous ecosystem, the Metaverse has democratic properties, allowing participants to participate in maintaining order and functioning properly. In Decentraland, users through a Decentralized Autonomous Organization (DAO) \cite{hassan2021decentralized} can propose and vote on policies created.  \\ \hline
			
			4 & Cultural heritage protection & The Metaverse can be used to rebuild damaged or inaccessible heritage buildings \cite{allam2022metaverse}. Computers scan the raw fragments and use machine learning algorithms to predict their original configuration. Throughout the process, artificial intelligence is guided by humans to ensure accurate reconstruction of the fragments. At the same time, the digital cultural heritage formed by scanning will help record history, build digital libraries for researchers and education, and enrich cultural vitality \cite{buhalis2022mixed,gaafar2021metaverse}. In addition, the virtual environment of the Metaverse and fully immersive virtual reality can be used to track the situation of cultural heritage on a regular basis, thus allowing it to be preserved in a timely manner. \\ 
			\hline
			
			5 & Military & The immersive training brought about by the Metaverse can serve as an important military application \cite{nalbant2021computer}. AR-based synthetic training environments in virtual worlds can provide a realistic and convenient training experience \cite{upadhyay2022metaverse}. The simulation of multiple real-world combat environments allows soldiers to accomplish their training objectives while keeping them safe. The stream of data recorded in real time allows soldiers to perform specific training on enemy performance, improving training efficiency. \\ \hline
			
			6 & Virtual events & The Metaverse provides significant support for enabling improved integrated solutions through virtual events. With immersive experiences in virtual worlds, virtual event planners can host events that make participants feel like they are there. One of the most notable examples of a virtual event that showcases the potential of business opportunities is Fortnite\tablefootnote{https://www.epicgames.com/fortnite/en-US/home}. Anyone who buys a ticket to the virtual concert can attend without any physical or geographic restrictions. Thus, businesses can take advantage of the benefits of increased audience participation in virtual events through the virtual world. Moreover, Metaverse also offers better prospects for collecting audience data and evaluating audience behavior.  \\ \hline		
		\end{tabular}
	}
\end{table}	

\normalsize	

"Security" is defined as protection against unauthorized access to information assets and confidential data. The three core principles of security include protecting data confidentiality, maintaining data integrity, and facilitating data availability \cite{von2013information}. Privacy defines the ability to protect users' sensitive data. Security and privacy overlap a lot. While security controls can be satisfied without meeting privacy concerns, it is impossible to address privacy concerns without first adopting effective security practices. To some extent, privacy is achieved through security measures, and the mechanisms used to ensure data privacy are part of the data security strategies. Therefore, security and privacy need to be considered throughout the development of the Metaverse in order to ensure that the Metaverse operates safely. From the perspective of the elements in the Metaverse, security and privacy issues can be discussed from the following five aspects: data, network, privacy, content and application, Law and ethics issues.% as shown in Fig. \ref{fig:Security}.

\subsection{Data}
	
The Metaverse is a data world. The massive amounts of data acquired from IoT devices or generated by users or avatars contain a lot of useful but sensitive information from which meaningful data solutions can be derived. However, there are still serious unresolved issues about how data is captured and used, and the confidentiality, integrity, and availability of data are under threat. Therefore, data-related security issues are the primary concern.
	
\textbf{Data acquisition.} As the first step in data analysis, it is important to collect accurate and realistic data. The large amount of data collected in the Metaverse is generated by multiple heterogeneous sources, which poses a challenge to data quality. The source of the data and the reasonable determination of the integrity of the data are critical because the quality of the data affects almost all work in the Metaverse \cite{perez2018data}. For example, IoT devices such as wearable sensors and crowd sensing sensors may generate inaccurate data if not checked or faulty data, thus affecting the construction of reference items in the virtual world \cite{guo2017availability,su2020lvbs}. In addition to data input from IoT devices, a portion of the data comes from users. If users produce low-quality content for profit in UGC mode, it will mislead data analysis and model training related to user behavior in the Metaverse.
	
\textbf{Data storage.} There is a huge amount of data in the Metaverse that needs to be stored, and that data is constantly growing. The explosive growth of data puts forward higher requirements on storage capacity, expansion speed and data backup capability. Real-time interaction between users requires faster read and write speeds and higher security. At the same time, due to the diversity and complexity of user-generated content (UGC) data and their interrelations, the storage system needs to transform the model from a single file type to a variety of semi-structured and unstructured data relationship models. Due to the high infrastructure costs, these features make it difficult for centralized storage methods to keep up with the volume and type of data production. In addition, the risk of data leakage, tampering, or loss will be greatly increased. Therefore, centralized storage is no longer applicable to the Metaverse.
	
\textbf{Data-related attacks.} Most cyberattacks want to obtain or compromise sensitive data in a system without authorization, so data-related attacks are also one of the major security threats in the Metaverse. Common data attack methods include false data attacks and data tampering attacks. A false data attack is a form of data attack that affects the computing power of the control center when an attacker changes or modifies the original measurements provided by these sensors \cite{liang20162015}. False data attacks are mainly classified into three categories: false data injection (FDI) attack, replay attack, and zero dynamic attack \cite {cui2020detecting}. There are two main types of FDI attacks. An attacker can hack into the system and obtain configuration data, thereby manipulating the protected data in the system. Alternatively, an attacker can inject falsified data into the system as much as possible without being detected, posing a huge threat to the system and being difficult to identify in real time. In contrast to the FDI attack, the replay attack will repeatedly upload secret data to terminal devices over a set period of time. The attacker deceives the central authority by eavesdropping on data to send the same data over and over again. A zero-dynamic attack refers to using unstable zeros as the bug to attack the system \cite{hoehn2016detection}. In addition, many cyberattacks involve data tampering. To gain control of the system, attackers change the system configuration file by inserting new files that perform malicious activities. To hide the access records, deleting or modifying the log file is often used. 
	
\subsection{Network}
	
The network layer of the Metaverse is based on the traditional network architecture and combines wireless communication technology, so an attack on the traditional network is also a threat to the Metaverse \cite{tang2022roadmap}.
	
\begin{enumerate}
	\item 	\textbf{Single point of failure (SPoF) \cite{guo2008dcell}.} This failure is prone to occur when there are no redundant personnel, facilities, equipment, applications, or other resources \cite{guo2008dcell,yao2013cascading}. This is a potential risk caused by a flaw in the design, implementation, or configuration of the system where the failure of one of the nodes will bring the entire system to a halt. When a system has a SPoF failure, both productivity and security are compromised.
	
	\item  \textbf{Man in the middle attacks.} This kind of attack refers to the attacker intercepting the normal communication data and tampering and sniffing the data, but the communication parties do not know \cite{conti2016survey,ahmad2018man}. An attacker may simply eavesdrop on the conversation to obtain information such as user login credentials, private information, and so on. However, they may also simulate other users to manipulate the conversation. An attacker may send an error message or share a malicious link, which can cause a system crash or enable other cyber-attacks. Typically, legitimate users do not realize they are actually communicating with an illegal third party until after the damage has occurred. Common man in the middle attacks include DNS spoofing, IP spoofing, ARP spoofing, and Wi-Fi hacking.
	
	\item  \textbf{Distributed denial-of-service (DDoS) attacks \cite{peng2007survey}.} DDoS attack is a computer (or computers) that prevents a server from performing its tasks. Attackers create botnets and large numbers of infected devices to direct fake traffic onto a network or server \cite{nazario2008ddos,vishwakarma2020survey}. Attackers overwhelm servers by flooding them with requests, forcing sites, servers, and applications to shut down. DDoS attacks can cause websites to crash, fault, or load slowly. DDoS attacks can occur at the network level, such as by sending large numbers of SYN/ACC packets that can overwhelm a server, or at the application level, such as by executing complex SQL queries that crash a database. SQL injection attacks use SQL queries to inject malicious code into a site or application to exploit security vulnerabilities and obtain or corrupt private data.
\end{enumerate}
		
\subsection{Privacy} 
	
As users engage in virtual activities and have more immersive experiences in virtual worlds, they will create huge amounts of data. At the same time, the Metaverse collects a large amount of user-related data from different sensors (e.g., wearables, microphones, hearts, user interactions). This will be of great public concern because such a large collection of data could easily lead to privacy violations \cite{di2021metaverse}. Since the Metaverse is a mapping of the real world, the user data in the virtual world will correspond to the identity information in the real world. Once the privacy data is leaked, it means that a large amount of sensitive information related to the biometric characteristics and behavior of the user is lost, causing great privacy risks to users. As users transfer more of their personal data into the Metaverse, the risk of sensitive or confidential data leakage will increase. Meanwhile, the implementation of the Metaverse requires the use of more sensors in users' living environments. Real-time monitoring of user behavior and physiological data will make these devices increasingly vulnerable to targeted cyberattacks. In this respect, the Metaverse has serious privacy implications. Online information leakage will provide opportunities for fraud, stalking, and other illegal activities in the real world \cite{falchuk2018social}. In addition, when data about user behavior is transformed into digital assets, it poses a serious threat to data management \cite{tang2022roadmap}. The ownership of digital assets is an issue worth discussing. In addition to monetizing this data, there are many problems to be solved in the storage, processing, and protection of private data.
	
\subsection{Content and applications}
	
\subsubsection{Identity} In the Metaverse, user identity is a very important credential, but it is also vulnerable to attack. The management and authentication of user identities are the common challenges \cite{wang2022survey}. VR, one of the supporting technologies of the Metaverse, is a vulnerable target for identity theft in virtual worlds. By gaining access to the motion tracking data of the user's VR headset, the attacker generates a digital copy to impersonate the user. At the same time, user network credentials are also the target of attackers. As long as access to a user's network credential is obtained, the attacker can log in to the user's account at any time to perform illegal activities. Moreover, the attacker may have access to personal information stored in Metaverse user profiles. Thus, it is necessary to provide more personal data when performing personal verification in the Metaverse \cite{tang2022roadmap}. While existing concepts of digital identity often remain in closed systems (each system has its own set of identities), new technologies such as NFTs indicate a new shift in digital identity. Unfortunately, the NFT is still in its early stages, and different attacks have been reported. Illegals can steal a person's digital identity by exploiting the weakest link in the chain, namely humans. In particular, they can carry out phishing, social engineering attacks, and scams. Therefore, without an appropriate metaverse identity management solution, it is impossible for users to identify and trust each other.
	
\subsubsection{Economy} The virtual economy is an important part of the Metaverse \cite{yang2022fusing}. Users experience, create, and communicate in the Metaverse with the goal of cash realizable value. The virtual human is the subject of the Metaverse digital economy. The value creation of virtual human economic behavior can realize the value-added of virtual original value. As the foundation of the Metaverse economy, cryptocurrencies, NFTs, and other digital assets are supported by most metaverse platforms, and these are likely to become the dominant form of Metaverse value exchange. However, new ways of operating bring skill and trust challenges for users and managers. Regulators may lack the authority and effective management of these transactions. Meanwhile, the storage and management of digital assets depend on the reliability of the system, such as encrypted storage and smart contract-based management and exchange. Criminals may exploit vulnerabilities in the system to steal digital assets.
	
\subsubsection{Society and the individual} The Metaverse integrates all kinds of devices in the real world, and the virtual world and the real world influence each other. If the security vulnerabilities in the virtual world are exploited by hackers, it will cause serious harm to the real world. For example, hackers attack social infrastructure such as smart grids to profit from it \cite{liang2016review, musleh2019survey}. The side effect of the highly immersive experience of the Metaverse is that the user becomes less aware of the physical space. The immersive and highly engaging experience that XR enables is the most prominent feature of virtual worlds. However, VR also isolates individuals from the real world for certain periods of time. The user immersed in a VR experience has no audio-visual connection to the real world. Therefore, the security issues in the virtual world caused by virtual reality also extend to the physical security issues in the user environment. If indoor sensors are hacked, private information such as the user's location will be leaked, which is a big security risk \cite{casey2019immersive}. In addition, being immersed in a virtual world for a long time can have a negative impact on one's health. With the continuous development of interaction methods, people have more and more convenient access to the Internet, and people spend more and more time on it, which greatly increases the probability of people suffering from Internet syndrome \cite{spiegel2018ethics}.

\subsection{Law and ethics}
	
The Metaverse provides people with a freer environment, followed by more complex social relations. The moral and legal rules of the real world no longer apply to the Metaverse. The first is a discussion of moral issues. The ethical and moral problems of the Metaverse refer to the conflict between the moral norms of the Metaverse and the real world \cite{ning2021survey}. As a next-generation network, the Metaverse must contain rules that can constrain user behaviors. With the development of metaverse interaction technology, many undiscussed moral issues will emerge, and the original moral and ethical norms will no longer play a role. If new rules are not set in time, the development of the Metaverse will be inhibited. The second is the legislation. The Metaverse is closely linked to the real world and constantly affects society. The new generation network is under construction, and the Metaverse must fully consider data privacy protection. With the rapid development of the digital environment, the current regulations will need to be constantly updated. The development of clearer guidance ensures that full consent is obtained prior to the use of collected data. As social media experiences become more immersive, concerns around data collection and use will inevitably put pressure on legislative reform.
	
\subsection{Security countermeasures}
	
The stable operation of the Metaverse requires safe and reliable systems. Existing problems with the original Internet and new problems arising from the convergence of technologies in the Metaverse will need to be solved. Based on these security threats, we summarize common security countermeasures.

\subsubsection{Data} The amount of data in the Metaverse is huge, and the original method of data management no longer works. Big data technology can be used to develop a new way of data management based on the Metaverse \cite{cai2022compute,han2022dynamic,ooi2022sense,sun2022matrix}. Big data technology enables the storage, processing, and analysis of massive amounts of data in the Metaverse to gain valuable insights from the data. Data security should be considered in the Metaverse, and it aims to protect vulnerable data from loss or theft. Sensitive information, such as user biometric and behavioral data, is the most vulnerable. Therefore, it is necessary to provide some mechanism to protect private business and personal data in the Metaverse. Firstly, ensure the physical security of the servers and user devices \cite{mousa2020database}. Whether a user's data is stored locally, in an enterprise data center, or in a public cloud, managers need to ensure that the facility is not compromised or damaged. Secondly, reasonable access management and control are very necessary. The principle of minimum access rights \cite{singh2019managing,guclu2020new} should be followed throughout the Metaverse. For example, access to databases and networks should be given minimal permissions for each legal action. This is to protect data and functions from errors or malicious acts. Finally, maintaining a usable and well-tested backup copy of all critical data is a core component of any strong data security policy. All backups should have the same security measures in place to control access to the primary database and the core system \cite{vidal2022new}.
	
\subsubsection{Network} Network security is critical to protecting customer data and information, keeping shared data secure, ensuring reliable access and network performance, and preventing cyber threats. Network security programs prevent the leakage of sensitive data, which helps companies gain users' trust and avoid additional losses. The stable operation of network systems and the legal access of data are the preconditions of providing good service to users. The existence of massive heterogeneous devices in the Metaverse brings new challenges to network security \cite{tang2022roadmap}. Due to the limitations of small devices, the Metaverse is better suited for embedded trust and security. At the same time, the zero trust model is also a security measure that can be considered in the Metaverse. The zero trust model requires strict identity checks and also uses continuous authentication and verification to ensure that illegal visitors are kept out or severely restricted \cite{ahmed2020protection,mehraj2020establishing}. Due to the large number of datasets to be hosted in the Metaverse, zero trust is the most effective way to reduce or eliminate the theft of sensitive information. Cloud computing and cloud storage are one of the supporting technologies of the Metaverse, so cloud security cannot be ignored \cite{huang2010new,chen2014collaborative,ahmed2014cloud,singh2017cloud}. Applications and workloads are moving to the cloud rather than relying entirely on local storage. Traditional networks need innovation to adapt to the transfer and access from local to cloud. Software defined network (SDN) \cite{hu2014survey,bhushan2019distributed} and Software defined wide area network (SD-WAN) support network security solutions in private cloud, public cloud, and hybrid cloud firewall as a service (FWaaS) deployments \cite{arins2015firewall,zeineddine2018stateful}.
	
\subsubsection{Privacy} 	Encryption and anonymization of user data are the basis of privacy protection in the Metaverse. The Metaverse needs to provide users with an easy-to-use data management platform that allows them to control what is shared. Transparency is essential to data privacy, which means accurately identifying the data to be collected. Since data in the Metaverse is likely to come from a range of sources, it is also important to be able to pinpoint the origin of each data element. The XR captures the user's biometric data, facial information, etc., or from the virtual world itself, which must be specially protected. Thereby, a digital biometrics based ID powered by blockchain becomes the solution \cite{delgado2019blockchain}. Biometric data can be used as a cryptographic basis to generate a pair of public and private keys. These keys will act as proof of identity on the network, enabling their holders to sign and receive transactions. Digital IDs based on key pairs can provide a more secure and protected identity. In addition, UGC will become an important part of the Metaverse. But UGC contains a lot of sensitive user information, which has a high privacy risk. Metaverse can develop blockchain-based privacy protection frameworks and schemes to regulate users' behavior to prevent their privacy leakage \cite{song2018personal, zhang2022security}.
	
\subsubsection{Content and application} 	In the Metaverse, the identity of the user and the Metaverse economy require a safe and stable operating environment and reasonable rules, and appropriate measures need to be discussed. The first is user identity management in the Metaverse. Users can have multiple completely different identities and can freely migrate between different virtual worlds. The digital identity of the user must be persistent. It cannot be copied, modified, or deleted by any individual or institution. It also implies uniqueness. A person has only one identity in a system, and each system identity corresponds to only one person. Moreover, the system should always take measures to protect the user's identity, such as blockchain-based identity protection \cite{jacobovitz2016blockchain}, to ensure that access and use of personal information without consent is strictly prohibited. Identity and behavior are the basis of user identification in the Metaverse. AI technology can provide automatic authentication and user behavior analysis \cite{adams2022virtual, zhang2022artificial}. Digital assets such as avatars and skins in social media are mostly managed by platforms. The Metaverse should develop a self-sovereign identity system to realize identity information based on fine-grained control, where users control the composition and access of their identity \cite{muhle2018Survey, ferdous2019search}. Second is the economic management of the Metaverse. The creator economy dominated by UGC is an important part of the Metaverse \cite{wang2022survey}. Unlike traditional e-commerce, the Metaverse economy is more liberalized and open. Blockchain and the development of emerging industries are the promotion of economic development \cite{ning2021survey}. Blockchain of a decentralized autonomous organization (DAO) can prevent the risk of centralized organization \cite{beck2018governance,liu2021technology}. Since blockchain uses proof-of-work as a consensus mechanism, it is more secure and more suitable for e-commerce platforms than traditional methods. The ownership of digital assets is the guarantee of the stability of the virtual economy system, which is directly related to the meta-universe economy. Blockchain stores submitted transactions to enable tracking and protection of digital assets in the Metaverse. These transactions or records are stored as blocks, and each block is encrypted and hashed to ensure ledger invariance and security \cite{gadekallu2022blockchain,huckle2016internet}. To summarize, the potential security countermeasures from content and applications should be addressed successfully.

\section{Opportunities}  \label{sec:Trends}
	
\subsection{Open challenges}
	
\subsubsection{The challenge of data.} The first problem is the data size. It could be 100 or 1,000 times more than it is now. In fact, the growth of data may be greater than this value. The main reason is that while more modes of interaction and more terminal growth are to be expected, it is the new applications and modes that new technologies will bring that are hard to fully anticipate. For example, the explosive application growth brought by mobile Internet technology has generated a large amount of UGC data, which is far larger than the total historical data in the fixed network era. Today, new applications are already generating huge amounts of data, such as autonomous driving, which has the potential to generate 100 terabyte (TB) of data per car per day. If intelligent vehicles become the terminal for individuals to enter the Metaverse, full interaction will also raise the volume of data to another level. The volume of data generated by these unpredictable new models is the main driver of growth. The second issue is data processing capacity building. The Metaverse is a world of data, and the ability to process data is a problem that must be overcome. First, data storage and access issues need to be addressed. This is not only a large number of users from any location accessing, but also a large number of various system accessing and data exchange requests. At the same time, the protection of data rights and security in the process of access and use, the effective protection of data property rights, and the prevention of data abuse are great challenges to data processing ability. In addition, the timely response ability of data processing is also very important. Very long delays are unable to meet the immersive design of the Metaverse. The indispensable basic technologies such as edge computing, distributed data, and blockchain that have been applied so far will be used again in the construction of the Metaverse to help realize the storage, access, right confirmation, openness, and circulation of data, so as to maintain the sustainable operation of the Metaverse system.
	
\subsubsection{Computing power.} Computing power refers to the ability to process data and is related to the calculation, storage, and transmission of data. Computing power is a new production factor and the cornerstone of building a digital society. The construction of Metaverse, a new digital environment, needs the support of super computing power. The Metaverse is the integration of many information technologies, including AI technology, input and output technology, blockchain technology, etc. It is also the fusion of virtual and real space. Therefore, the Metaverse puts forward very high requirements on information and communication technology and related infrastructure, and it is impossible to realize the Metaverse without the strong support of a powerful computing network. The development of the Metaverse still faces many challenges, not only technological challenges but also energy challenges. The large-scale data visualization and large-scale computing brought by the Metaverse require not only computing power, but also electricity.
	
\subsubsection{Interaction techniques.} As a medium between the virtual world and the real world, common Metaverse interactive technologies include XR technology, somatosensory technology, and brain-computer interface. However, the current interaction technology cannot meet the requirements for portability and transparency. Future interaction technology should not only strive to achieve lightweight, wearable, and convenient interactive devices, but also try to realize the transparency of the media and improve the user's immersion experience. At present, the cost of interactive equipment in somatosensory technology and XR technology is high, and it is difficult to popularize. AR uses lightweight equipment for short-term use, while VR requires heavy and expensive equipment for long-term use. Some methods integrate the advantages of AR and VR and use hardware switching to achieve the goal, but it is more expensive and inconvenient than a single device. In addition, holograms are not widely used in the Metaverse, but they have potential. BCI has safety risks, such as high surgical risk.
	
\subsubsection{Interdisciplinary integration.} The influencing factors in the real world are complex and changeable, and involve the research of many disciplines. The Metaverse is a virtual mapping of the real world, so interdisciplinary research of the Metaverse is necessary. The Metaverse involves all aspects of people's lives, such as medicine, education, the economy, and so on. It needs to make use of relevant theoretical knowledge and research to improve its own construction. The generation and maintenance of virtual currency in the Metaverse are closely related to the real economic system, and the integration and development of the two will have a profound impact on the real world. Metaverse also uses psychology and medicine, combined with AI and other machine learning technologies, to deeply analyze and understand user behavior, providing users with a more immersive experience or for use in psychotherapy, education, training, and other industries.
	
\subsection{Future direction}
	
The Metaverse is a hotly debated emerging Internet concept and a collection of cutting-edge technologies. Achieving the functions of the Metaverse requires the maturity and application of these technologies. For example, the Metaverse needs artificial intelligence technology to supervise information; otherwise it cannot control the spread of harmful information; it needs a mature blockchain mechanism to store and authenticate identities; it needs data computing to complete data analysis, which involves cloud computing and edge computing; a 5G or 6G network is required to complete data transmission; the display of the Metaverse requires virtual reality technology, etc. In addition to the above technologies, the Metaverse also needs other technologies such as 3D rendering, XR, BCI, wearable devices, robotics, etc. Although the deployment of these technologies is not yet in place, they are developing rapidly. By integrating a number of cutting-edge information technologies, Metaverse attempts to create a more immersive and open digital space to promote the increase of Internet penetration and the upgrade of the digital economy. This model is likely to become the mainstream development model of the future digital economy. At present, this concept is still in its infancy, and once it matures and develops, it will have a broad impact on the international economy, politics, and society, as well as on the international landscape and the relationships between countries.
	
\section{Conclusion} \label{sec:conclusion}
	
In this article, we review the latest research results related to the Metaverse. Several existing surveys have investigated some aspects of the Metaverse, but they focus only on technology or security, and lack a comprehensive review of the concepts and applications of the Metaverse. As far as we know, this is the first paper to review the basic concepts, technologies, applications, security issues, and security solutions of the Metaverse in detail. Due to the broad prospects of Metaverse and other emerging technologies, a great deal of research based on these related technologies has been carried out in the past two years. Therefore, we first review the literature on the Metaverse and its applications, as well as the literature that provides solutions to security and privacy issues. We summarize the technical framework and application status of the Metaverse. More specifically, we focus on current security threats and solutions to the Metaverse, and propose the challenges and future directions for the Metaverse.

\section*{Acknowledgment}
	
This research was supported in part by the National Natural Science Foundation of China (Grant Nos. 62002136 and 62272196), Natural Science Foundation of Guangdong Province (Grant No. 2022A1515011861), Guangzhou Basic and Applied Basic Research Foundation (Grant No. 202102020277), and the Young Scholar Program of Pazhou Lab (Grant No. PZL2021KF0023).

\bibliographystyle{ACM-Reference-Format}
\bibliography{surveyMetaverse}
	
\end{document}